\documentclass[aps,prd,eqsecnum,amssymb,nofootinbib,floatfix,showpacs,a4paper]{revtex4-2}

\usepackage{amsmath}
\usepackage{amssymb}
\usepackage{mathtools}
\usepackage{dcolumn}
\usepackage{graphicx}
\usepackage{xcolor}
\usepackage{multirow}
\usepackage{enumerate}
\usepackage{nicematrix}

\usepackage[utf8]{inputenc}
\usepackage[T1]{fontenc}

\newcommand{\bdm}{\begin{displaymath}}
\newcommand{\edm}{\end{displaymath}}
\newcommand{\be}{\begin{equation}}
\newcommand{\ee}{\end{equation}}
\newcommand{\bea}{\begin{eqnarray}}
\newcommand{\eea}{\end{eqnarray}}
\newcommand{\ben}{\begin{eqnarray*}}
\newcommand{\een}{\end{eqnarray*}}
\newcommand{\bse}{\begin{subequations}}
\newcommand{\ese}{\end{subequations}}
\newcommand{\ba}{\begin{array}}
\newcommand{\ea}{\end{array}}

\newcommand{\btau}{\boldsymbol{\tau}}
\newcommand{\btheta}{\boldsymbol{\theta}}
\newcommand{\bxi}{\boldsymbol{\xi}}

\newcommand{\Ca}{C_\textrm{a}}

\newcommand{\chii}{\chi_\text{c}}
\newcommand{\cmin}{C_\textrm{min}}

\newcommand{\F}{\mathcal{F}}

\newcommand{\fc}{f_\text{c}}

\newcommand{\dlangle}{\langle\!\langle}
\newcommand{\drangle}{\rangle\!\rangle}
\newcommand{\avii}[1]{\Big\langle\!\!\Big\langle#1\Big\rangle\!\!\Big\rangle}

\newcommand{\md}{{\mathrm{d}}}

\newcommand{\To}{T_\text{o}}
\newcommand{\ti}{t_\text{c}}

\begin{document}

\title{Banks of templates
for directed and all-sky narrow-band searches of continuous gravitational waves
from~spinning~neutron~stars with several spindowns}

\author{Andrzej Pisarski}
\email{a.pisarski@uwb.edu.pl}
\author{Piotr Jaranowski}
\email{p.jaranowski@uwb.edu.pl}
\affiliation{Faculty of Physics,
University of Bia{\l}ystok,
K.\ Cio{\l}kowskiego 1L, 15--245 Bia{\l}ystok, Poland}

\date{\today}

\begin{abstract}

We construct efficient banks of templates suitable for searches of continuous gravitational waves
from isolated spinning neutron stars. We assume that the search algorithm is based on the time-domain
maximum-likelihood $\F$-statistic and we consider narrow-band searches with several spindown parameters included.
Our template banks are suitable for both all-sky and directed searches
and they enable the usage of the fast Fourier transform in the computation of the $\F$-statistic
as well as, in the case of all-sky searches, efficient resampling of data to barycentric time.

\end{abstract}


\maketitle

\section{Introduction}

In the years 2015--2020 second-generation ground-based gravitational-wave (GW) detectors
Advanced LIGO \cite{aLIGO} and Advanced Virgo \cite{aVirgo}
completed three observational runs O1, O2, and O3,
during which ninety GW transient signals from coalescence of pairs of black holes or neutron stars were registered
\cite{GWTC1,GWTC2,GWTC2.1,GWTC3}.
Many other possible astrophysical sources of gravitational waves remain undetected,
among them long-lasting almost monochromatic gravitational waves
coming from single rotating neutron stars located in our Galaxy
(sources of continuous gravitational waves were recently reviewed in \cite{2015Lasky,2018GG,2019SB}).
There exist several different data-analysis pipelines developed to search
for this kind of continuous GW signals \cite{2021TKS}.
Our paper discusses construction of efficient bank of templates
needed to perform searches using the time-domain $\F$-statistic pipeline.
The $\F$-statistic introduced in \cite{98JKS} is a detection statistic
that follows from the maximum-likelihood principle of detection of deterministic signals buried in detector's noise
(the maximum-likelihood detection in Gaussian noise
is described e.g.\ in \cite{W71,JKlrr,JKbook}).
Implementation of the maximum-likelihood approach to all-sky narrow-band searches of almost monochromatic
GW signals were done in details in the series of articles \cite{98JKS,JK99,JK00,ABJK02,ABJPK10}.

In the case of all-sky searches (in which neither the location of the source in the sky nor its GW frequency is known),
the time-domain $\F$-statistic pipeline was first used in analysis of Virgo VSR1 data \cite{2014LSC/VC-CQGb},
and then also for LIGO/Virgo O1 \cite{2017LSC/VC-PRDa,2018LSC/VC-PRDa},
O2 \cite{2019LSC/VC-PRDa}, and O3 \cite{2022LSC/VC-arXiv} data.
In the search of VSR1 data \cite{2014LSC/VC-CQGb}, bank of templates build in section IV of \cite{ABJPK10} was used,
whereas all-sky searches of O1--O3 data \cite{2017LSC/VC-PRDa,2018LSC/VC-PRDa,2019LSC/VC-PRDa,2022LSC/VC-arXiv}
employed banks of templates constructed in Ref.\ \cite{2015PJ}.
Efficient banks of templates for $\F$-statistic-based directed searches
(in which the position of the source in the sky is known, but its GW frequency is not)
were considered in \cite{2011PJP}.
Our paper generalizes and unifies the results of both Refs.\ \cite{2011PJP} and \cite{2015PJ}.
In Ref.\ \cite{2011PJP}, banks of templates in two-dimensional parameter space
(spanned by GW frequency and the first spindown parameter) were considered,
they were suitable only for directed searches. Reference \cite{2015PJ} constructed banks of templates
suitable for all-sky searches in 4-dimensional parameter space (spanned by GW frequency, the first spindown,
and two more parameters related to the unknown position of the source in the sky).
The procedure of construction of template banks presented in section \ref{secBofT} of this work
can be used both for directed and all-sky searches, and for any number of spindown parameters included.
We have implemented this procedure in a computer program that is publicly available
(see section \ref{secDiscussion} below for more details),
and we have tested it for parameter spaces with dimensions from two to six (inclusive).
All our template banks enable the usage of the fast Fourier transform algorithm in the computation of the $\F$-statistic
and, in the case of all-sky searches, efficient resampling of data to barycentric time.
This work completes the research on the construction of banks of templates developed in \cite{2011PJP,2015PJ}.
We expect that the bank of templates constructed in the current work
will be employed in the near future to analyze data from the LIGO/Virgo/KAGRA detectors \cite{aLIGO,aVirgo,KAGRA}
using the time-domain $\F$-statistic pipeline.

As is explained in section \ref{secBofT} below,
our approach to template placement is based on usage of isoheights of the signal-free autocovariance function of the $\F$-statistic.
Another common approach in constructing bank of templates for GW searches
is based on the notion of a metric in the space of signal's parameters.
This approach was initiated in the works \cite{1996BSD,1996Owen} for searches of gravitational waves from coalescence of compact binaries.
For continuous-wave searches it was developed in \cite{2007Prix}
and then expanded in many papers (see, e.g., \cite{2013WettePrix,2014Wette,2015Wette,2016Wette,2022W4}).
In Appendix A of \cite{2015PJ}, the formalisms based on the notions of autocovariance function of the $\F$-statistic
and metric in the space of signal's parameters are compared to each other.
It is shown there that for large signal-to-noise ratios,
the autocovariance function plays the role of the \emph{match} introduced in the metric approach.

The match function is usually expanded in a power series with respect to, assuming small,
separations between the parameters of the template and the signal.
Allen in \cite{2019Allen} proposed another 'spherical ansatz' for larger separations.
We use the standard approach in our work,
i.e.\ we assume that we are dealing with small enough separations.
Also Allen in \cite{2021Allen} (see also \cite{2021AllenShoom}) proposes to distinguish,
when looking for optimal banks of templates,
between setting the best possible upper limits on unknown signal's parameters
and maximizing the expected number of detections.
We comment on this approach in section \ref{secDiscussion}.

The article is organized as follows:
in section \ref{secFstat} we introduce the simplified model of the GW signal [given by Eqs.\ \eqref{hPhi}]
and compute the $\F$-statistic for that signal together with
the autocovariance function of the $\F$-statistic in the case when there is no signal in the data.
Finally, in a situation where the parameters of the template differ only slightly from the parameters of the signal present in the data,
we show that the signal-free autocovariance function can be expressed in terms of the reduced Fisher matrix.
In section \ref{SecRFmatrices}, we explicitly compute the reduced Fisher matrices needed
for directed searches with up to five spindown parameters included
and for all-sky searches with up to three spindown parameters.
We also show how to combine the Fisher matrices evaluated for individual detectors
to get the reduced Fisher matrix for the whole network of detectors.
Section \ref{secBofT} is devoted to the details of the construction of banks of templates
suitable for both all-sky and directed searches, and for different number of spindown parameters included.
We discuss our results in section \ref{secDiscussion}.
In appendix \ref{appendixAdstar} we provide some useful information about the $A_d^*$ lattices,
while appendix \ref{AS2-2D} is devoted to the details of the construction of the 2-dimensional $S_2$ grid.

\section{Signal-free autocovariance function of the $\F$-statistic}
\label{secFstat}

For the convenience of the reader, we present in this section
a short derivation of the signal-free autocovariance function of the $\F$-statistic,
which we use to construct bank of templates (more details can be found in section 2 of \cite{2015PJ}).
We consider \emph{narrow-band} searches for \textit{almost monochromatic} GW signals.
The searched signal is almost monochromatic in the sense that
the modulus of the Fourier transform of the signal's waveform
is well concentrated (for frequencies $f\ge0$) around some `central' frequency $\fc$ within the bandwith of the search.
The search is also narrow-band, which means that the frequency bandwith of the search
is small enough to assume that the spectral density $S_{n}$ of detector's noise
can be replaced by a constant $S_0:=S_{n}(\fc)$ within the bandwith.
We also assume that the noise $n$ in the detector is an additive, stationary, Gaussian,
and zero-mean continuous stochastic process.
Then the logarithm of the likelihood function is approximately given by
(see, e.g., section 2 of \cite{2015PJ} for more details)
\be
\label{004}
\ln\Lambda[x] \cong \frac{2\,\To}{S_0}\,
\Big(\langle x\,h \rangle -\frac{1}{2}\langle h^{2}
\rangle\Big),
\ee
where $x$ denotes the data from the detector,
$h$ is the deterministic signal we are looking for in the data,
$\langle\,\cdot\,\rangle$ is the time averaging operator defined by
\be
\label{003}
\langle h \rangle := \frac{1}{\To}
\int\limits_{\ti-\To/2}^{\ti+\To/2}h(t)\,\md t.
\ee
Here $[\ti-\To/2;\,\ti+\To/2]$ denotes observational interval,
so $\ti$ is the instant of time in the middle of this interval
and $\To$ is the length of observation time.

We employ a simplified \emph{linear}\footnote{The linearity of the model means
that the phase $\Phi(t;\bxi)$ of the signal $h$ from Eq.\ \eqref{h}
is a linear function of the parameters $\bxi$, see Eq.\ \eqref{Phi}.
The details of the model are described in section V~B of Ref.\ \cite{JK99}.}
model of the GW signal from a rotating neutron star,
which can be written as
\begin{subequations}
\label{hPhi}
\begin{align}
\label{h}
h(t;h_0,\Phi_0,\bxi)
&= h_{0}\,\sin\big(\Phi(t;\bxi)+\Phi_{0}\big),
\\[1ex]
\label{Phi}
\Phi(t;\bxi) &= \sum_{k=0}^{s} \omega_k \Big(\frac{t}{\To}\Big)^{k+1}
+ \alpha_{1}\,\mu_{1}(t) + \alpha_{2}\,\mu_{2}(t).
\end{align}
\end{subequations}
Here $h_0$ and $\Phi_0$ are constant amplitude and initial phase of the signal, respectively,
and the vector $\bxi$ collects $s+3$ phase parameters,
\be
\label{vecxi}
\bxi = (\omega_0,\ldots,\omega_s,\alpha_1,\alpha_2).
\ee
The dimensionless frequency $\omega_0$ and spindown $\omega_k$ parameters are defined as
\be
\label{omegak}
\omega_0 := 2\pi f_0 \To,
\quad
\omega_k := \frac{2\pi}{(k+1)!} f_0^{(k)} \To^{k+1}
\ (k=1,\ldots,s),
\ee
where $f_0$ and $f_{0}^{(k)}$ ($k=1,\ldots,s$)
is respectively an instantaneous frequency and the $k$th time derivative of the instantaneous frequency
of gravitational wave at the solar system barycenter (SSB), evaluated at $t=0$.
The parameters $\alpha_1$ and $\alpha_2$ depend on the position of the GW source in the sky:
$\alpha_1 := 2\pi f_{0}(\sin\alpha\cos\delta\cos\varepsilon+\sin\delta\sin\varepsilon)$,
$\alpha_2 := 2\pi f_{0}\cos\alpha\cos\delta$,
where $\alpha$ is the right ascension and $\delta$ is the declination of the source,
$\varepsilon$ is the obliquity of the ecliptic.
The functions $\mu_1$ and $\mu_2$ are known functions of time,
which depend on the motion of the detector with respect to the SSB
(see section 2 of \cite{2015PJ} and section~II of Ref.\ \cite{98JKS} for more details).

Let us emphasize that we use the simplified model of the signal defined above
only to construct a grid of parameters for which we want to calculate the $\F$-statistic.
However, in the $\F$-statistic-based pipeline \cite{2014LSC/VC-CQGb,2017LSC/VC-PRDa,2018LSC/VC-PRDa,2019LSC/VC-PRDa,2022LSC/VC-arXiv},
to obtain a numerical value of the $\F$-statistic,
we use an exact, not an approximate, signal model.
In Ref.\ \cite{JK99}, it was verified that the linear model reproduces well the covariance matrix,
defined as the inverse of the Fisher matrix,
for the maximum-likelihood estimators of the exact GW signal, both for directed and all-sky searches.
Moreover, this reproduction is better the longer the observation time
(provided one uses large enough number of spindown parameters).
So, the linear model is a good approximation for computing the Fisher matrix,
and in the rest of this work we use the Fisher matrices to construct banks of templates.

After introducing new parameters $h_{1}:=h_{0}\cos{\Phi_{0}}$ and $h_{2}:=h_{0}\sin\Phi_0$
(so that $h_0=\sqrt{h_1^2+h_2^2}$), the signal \eqref{hPhi} can be written as
\be
\label{22}
h(t;h_1,h_2,\bxi) = h_{1}\sin\Phi(t;\bxi) + h_{2}\cos\Phi(t;\bxi).
\ee
For observations which last at least several hours
and for GW frequencies of the order of tens of Hertz or higher,
to a good approximation $\langle\sin2\Phi\rangle\cong0$ and $\langle\cos2\Phi\rangle\cong0$.
Then the time average $\langle{h^2}\rangle$ approximately equals
\be
\label{25}
\langle h^{2} \rangle \cong \frac{1}{2}(h_{1}^{2} + h_{2}^{2}).
\ee
Substituting Eqs.\ \eqref{22} and \eqref{25} into \eqref{004}
we get the following formula for the log likelihood ratio of the signal \eqref{22}:
\begin{align}
\label{26}
\ln\Lambda[x;h_{1},h_{2},\bxi] &\cong
\frac{2\,\To}{S_0}\,\bigg(h_{1}\,\langle x(t)\sin\Phi(t;\bxi)\rangle
+ h_{2}\,\langle x(t)\,\cos\Phi(t;\bxi)\rangle-\frac{1}{4}(h_{1}^{2}+h_{2}^{2})\bigg).
\end{align}
We also compute, making use of \eqref{25}, the optimal signal-to-noise ratio
$\rho$ for the signal \eqref{22}:\footnote{The scalar product $(h_1|h_2)$ between waveforms
is defined e.g.\ in Eq.\ (2.2) of \cite{2015PJ}; we use here the approximation
$(h_1|h_2)\cong(2\To/S_0)\langle h_1h_2\rangle$, see section 2 of \cite{2015PJ} for its justification.}
\be
\label{snr}
\rho(h_0) = \sqrt{(h|h)}
\cong \sqrt{\frac{2\To}{S_0} \langle h^{2} \rangle}
\cong h_0 \sqrt{\frac{\To}{S_0}}.
\ee

In the next step we maximize $\ln\Lambda$ with respect to the parameters $h_{1}$ and $h_{2}$.
To do this we solve equations $\partial\ln\Lambda[x;h_{1},h_{2},\bxi]/\partial h_{i}=0$ ($i=1,2$).
Their unique solution $\widehat{h}_{1}\cong2\langle x\sin\Phi\rangle$ and
$\widehat{h}_{2}\cong2\langle x\cos\Phi\rangle$
gives the maximum-likelihood estimators of the parameters $h_{1}$ and $h_{2}$.
To get the \emph{$\F$-statistic} we replace in \eqref{26}
the parameters $h_{1}$ and $h_{2}$ by their estimators $\widehat{h}_{1}$ and $\widehat{h}_{2}$:
\begin{align}
\label{29}
\F[x;\bxi] := \ln \Lambda[x;\hat{h}_{1},\hat{h}_{2},\bxi]
\cong \frac{2\,\To}{S_0}
\bigg(\langle x(t)\,\sin\Phi(t;\bxi)\rangle^{2}+\langle x(t)\,\cos\Phi(t;\bxi)\rangle^{2}\bigg).
\end{align}

To construct efficient banks of templates we consider the expectation value\footnote{
The subscript `1' means computation in the case when the data contains some GW signal.}
of the $\F$-statistic \eqref{29} in the case when the data $x$ contains some GW signal $h$,
i.e.\ when $x(t) = n(t) + h(t;\btheta')$,
where $\btheta'=(h_1',h_2',\bxi')$ collects the parameters of the signal present in the data.
This expectation value can be written as
\be
\label{EF3}
\text{E}_1\{\F[x;\bxi]\}:=\text{E}\big\{\F\big[n(t) + h(t;\btheta');\bxi\big]\big\}
\cong 1 + \frac{1}{2} \, \rho(h_0')^2 \, C_0(\bxi,\bxi'),
\ee
where $\rho(h_0')$ is the signal-to-noise ratio \eqref{snr}
computed for the signal $h(t;\btheta')$ (so $h_0'=\sqrt{h_1'^2+h_2'^2}$)
and $C_0(\bxi,\bxi')$ is the \emph{autocovariance function} of the signal-free $\F$-statistic.\footnote{
In the signal-free case $x(t)=n(t)$ and the $\F$-statistic $\mathcal{F}[n;\bxi]$
is the \emph{random field} which depends on the parameters $\bxi$;
autocovariance function of $\mathcal{F}[n;\bxi]$ is defined as
$C_0(\bxi,\bxi') \coloneqq \mathrm{E}\big\{[\F[n;\bxi]-m_0(\bxi)][{\cal F}[n;\bxi']-m_0(\bxi')]\big\}$
(the subscript `0' means that the data contains only noise),
where $m_0$ is the signal-free expectation value of $\F$:
$m_0(\bxi) \coloneqq \mathrm{E}\{\F[n;\bxi]\}$.}
In section~IV of Ref.\ \cite{JK00} it was shown
that the autocovariance function $C_0$ for the narrow-band signal of the form \eqref{22}
can be approximated by
\begin{align}
\label{C02}
C_0(\bxi,\bxi') \cong \big\langle\sin\big[\Phi(t;\bxi)-\Phi(t;\bxi')\big]\big\rangle^{2}
+ \big\langle\cos\big[\Phi(t;\bxi)-\Phi(t;\bxi')\big]\big\rangle^{2},
\end{align}
Because the phase $\Phi$ of the signal \eqref{22} is a linear function of $\bxi$,
$C_0$ depends only on $\btau:=\bxi-\bxi'$,
so that one can rewrite Eq.\ \eqref{C02} as
\be
\label{C04}
C_0(\btau) \cong \langle\cos\Phi(t;\btau)\rangle^{2}
+ \langle\sin\Phi(t;\btau)\rangle^{2}.
\ee
$C_0$ attains its maximal value 1 for $\btau=\mathbf{0}$, i.e.\ for $\bxi=\bxi'$.

The final approximation we employ relies on assuming that $|\btau|\ll1$.
After expanding the right-hand side of \eqref{C04} in Taylor series around $\btau=\mathbf{0}$ up to terms quadratic in $\btau$,
we obtain the approximate autocovariance function $\Ca$:\footnote{We employ here the equalities: $\Phi(t;\btau=\mathbf{0})=0$
and $\partial^{2}\Phi(t;\btau)/\partial\tau_{k}\partial\tau_{l}=0$ ($k,l=1,\ldots,s+3$).}
\be
\label{Ca}
C_0(\btau) \cong \Ca(\btau) \coloneqq
1 - \sum_{k=1}^{s+3}\sum_{l=1}^{s+3} \tilde{\Gamma}_{kl}\,\tau_{k}\,\tau_{l},
\ee
where $\tilde{\Gamma}$ is the ($s+3$)-dimensional
\emph{reduced Fisher matrix} with elements equal to
\be
\label{Gamma}
\tilde{\Gamma}_{kl} \coloneqq \Big\langle\frac{\partial\Phi}{\partial\tau_{k}}
\frac{\partial\Phi}{\partial\tau_{l}}\Big\rangle -
\Big\langle\frac{\partial\Phi}{\partial\tau_{k}}
\Big\rangle \Big\langle\frac{\partial\Phi}{\partial\tau_{l}}
\Big\rangle, \quad k,l=1,\ldots,s+3.
\ee
The consequence of linearity of the signal's phase [see Eq.\ \eqref{Phi}]
is that the elements $\tilde{\Gamma}_{kl}$ \emph{do not depend} on the values of the parameters $\btau$.

\section{Reduced Fisher matrices}
\label{SecRFmatrices}

In this section we give explicit formulas for the reduced Fisher matrices defined in Eqs.\ \eqref{Gamma},
which are needed for constructions of template banks suitable for both directed and all-sky searches.
For convenience we replace the time $t$ by the dimensionless variable $\chi$
and introduce the dimensionless quantity $\chii$,
\be
\label{xchii}
\chi(t) \coloneqq \frac{t}{\To} - \chii,
\qquad
\chii \coloneqq \frac{\ti}{\To}.
\ee
Then the observational interval $t\in[\ti-\To/2;\,\ti+\To/2]$
is transformed into the interval $\chi\in[-1/2;1/2]$ of unit length.
We can also introduce averaging with respect to the variable $\chi$ defined as
\be
\dlangle g(\chi) \drangle \coloneqq \int\limits_{-1/2}^{1/2}g(\chi)\,\md\chi.
\ee
For any function $f$,
$\langle{f(t)}\rangle=\dlangle{f(t(\chi))}\drangle$,
where, according to Eq.\ \eqref{xchii}, $t(\chi)=(\chi+\chii)\To$.

\subsection{Directed searches with three spindown parameters}
\label{SubsecRFmD3}

To save space we first present the reduced Fisher matrix $\tilde{\Gamma}^{\text{dir}}_{s=3}$
for the case of directed searches with three spindown parameters included.
In this case the phase $\Phi$ of the signal is of the form [see Eq.\ \eqref{Phi}]
\be
\label{dir3phase}
\Phi(t;\btau) = \omega_{0}\frac{t}{\To} + \omega_{1}\bigg(\frac{t}{\To}\bigg)^2 + \omega_2\bigg(\frac{t}{\To}\bigg)^3
+ \omega_3\bigg(\frac{t}{\To}\bigg)^4,
\ee
where $\btau=(\omega_{0},\omega_{1},\omega_2,\omega_3)$.
The reduced Fisher matrix $\tilde{\Gamma}$ for the phase $\Phi$ defined above reads
(the matrix is symmetric, therefore we display below only its diagonal and upper-diagonal elements)
\be
\label{dir3mF}
\tilde{\Gamma}^{\text{dir}}_{s=3} = \begin{pmatrix}
  \frac{1}{12}
& \frac{1}{6}\chii
& \frac{1}{4}\chii^{2}+\frac{1}{80}
& \frac{1}{3}\chii^3+\frac{1}{20}\chii
\\[1ex]
\cdots
& \frac{1}{3}\chii^{2}+\frac{1}{180}
& \frac{1}{2}\chii^3+\frac{1}{24}\chii
& \frac{2}{3}\chii^4+\frac{2}{15}\chii^2+\frac{1}{840}
\\[1ex]
\cdots
& \cdots
& \frac{3}{4}\chii^4+\frac{1}{8}\chii^2+\frac{1}{448}
& \chii^5+\frac{3}{10}\chii^3+\frac{1}{80}\chii
\\[1ex]
\cdots
& \cdots
& \cdots
& \frac{4}{3}\chii^6+\frac{3}{5}\chii^4+\frac{1}{20}\chii^2+\frac{1}{3600}
\end{pmatrix}.
\ee
The elements of the matrix $\tilde{\Gamma}^{\text{dir}}_{s=3}$
depend only on the dimensionless parameter $\chii$.

\subsection{Directed searches with five spindown parameters}
\label{SubsecRFmD5}

In the case of directed searches with five spindown parameters included,
the phase $\Phi$ of the signal is of the form [see Eq.\ \eqref{Phi}]
\be
\label{dir5phase}
\Phi(t;\btau) = \omega_{0}\frac{t}{\To} + \omega_{1}\bigg(\frac{t}{\To}\bigg)^2 + \omega_2\bigg(\frac{t}{\To}\bigg)^3
+ \omega_3\bigg(\frac{t}{\To}\bigg)^4 + \omega_4\bigg(\frac{t}{\To}\bigg)^5 + \omega_5\bigg(\frac{t}{\To}\bigg)^6,
\ee
where $\btau=(\omega_{0},\omega_{1},\omega_2,\omega_3,\omega_4,\omega_5)$.
The reduced Fisher matrix $\tilde{\Gamma}^{\text{dir}}_{s=5}$ for the phase $\Phi$ defined above
is equal to (we display again only diagonal and upper-diagonal elements of the matrix)
\be
\label{dir5mF}
\tilde{\Gamma}^{\text{dir}}_{s=5}(\chii) = \begin{pNiceMatrix}
\Block{4-4}{\tilde{\Gamma}^{\text{dir}}_{s=3}(\chii)}
&&&
& \tilde{\Gamma}_{15}^{\text{dir}}(\chii)
& \tilde{\Gamma}_{16}^{\text{dir}}(\chii)
\\[1ex]
&&&
& \tilde{\Gamma}_{25}^{\text{dir}}(\chii)
& \tilde{\Gamma}_{26}^{\text{dir}}(\chii)
\\[1ex]
&&&
& \tilde{\Gamma}_{35}^{\text{dir}}(\chii)
& \tilde{\Gamma}_{36}^{\text{dir}}(\chii)
\\[1ex]
&&&
& \tilde{\Gamma}_{45}^{\text{dir}}(\chii)
& \tilde{\Gamma}_{46}^{\text{dir}}(\chii)
\\[1ex]
\cdots
& \cdots
& \cdots
& \cdots
& \tilde{\Gamma}_{55}^{\text{dir}}(\chii)
& \tilde{\Gamma}_{56}^{\text{dir}}(\chii)
\\[1ex]
\cdots
& \cdots
& \cdots
& \cdots
& \cdots
& \tilde{\Gamma}_{66}^{\text{dir}}(\chii)
\end{pNiceMatrix},
\ee
where $\tilde{\Gamma}^{\text{dir}}_{s=3}$ is given in Eq.\ \eqref{dir3mF} and where
\begin{subequations}
\begin{align}
\tilde{\Gamma}_{15}^{\text{dir}}(\chii) &= \frac{5}{12}\chii^4+\frac{1}{8}\chii^2+\frac{1}{448},
\\[1ex]
\tilde{\Gamma}_{16}^{\text{dir}}(\chii) &= \frac{1}{2}\chii^5 + \frac{1}{4}\chii^3 + \frac{3}{224}\chii,
\\[1ex]
\tilde{\Gamma}_{25}^{\text{dir}}(\chii) &= \frac{5}{6}\chii^5+\frac{11}{36}\chii^3+\frac{1}{96}\chii,
\\[1ex]
\tilde{\Gamma}_{26}^{\text{dir}}(\chii) &= \chii^6 + \frac{7}{12}\chii^4 + \frac{5}{112}\chii^2 + \frac{1}{4032},
\\[1ex]
\tilde{\Gamma}_{35}^{\text{dir}}(\chii) &= \frac{5}{4}\chii^6+\frac{29}{48}\chii^4+\frac{3}{64}\chii^2+\frac{1}{2304},
\\[1ex]
\tilde{\Gamma}_{36}^{\text{dir}}(\chii) &= \frac{3}{2}\chii^7 + \frac{43}{50}\chii^5 + \frac{31}{224}\chii^3 + \frac{3}{896}\chii,
\\[1ex]
\tilde{\Gamma}_{45}^{\text{dir}}(\chii) &= \frac{5}{3}\chii^7+\frac{13}{12}\chii^5+\frac{7}{48}\chii^3+\frac{1}{320}\chii,
\\[1ex]
\tilde{\Gamma}_{46}^{\text{dir}}(\chii) &= 2\chii^8+\frac{9}{5}\chii^6+\frac{5}{14}\chii^4+\frac{9}{560}\chii^2+\frac{3}{49280},
\\[1ex]
\tilde{\Gamma}_{55}^{\text{dir}}(\chii) &= \frac{25}{12}\chii^8+\frac{65}{36}\chii^6+\frac{35}{96}\chii^4+\frac{1}{64}\chii^2+\frac{1}{11264},
\\[1ex]
\tilde{\Gamma}_{56}^{\text{dir}}(\chii) &= \frac{5}{2}\chii^9+\frac{17}{6}\chii^7+\frac{89}{112}\chii^5+\frac{13}{224}\chii^3+\frac{3}{3584}\chii,
\\[1ex]
\tilde{\Gamma}_{66}^{\text{dir}}(\chii) &= 3\chii^{10}+\frac{17}{4}\chii^8+\frac{89}{56}\chii^6+\frac{39}{224}\chii^4+\frac{9}{1792}\chii^2+\frac{9}{652288}.
\end{align}
\end{subequations}

\subsection{All-sky searches with three spindown parameters}
\label{SubsecRFmAS}

In the case of all-sky searches with three spindown parameters included,
the phase $\Phi$ of the signal is of the form [see Eq.\ \eqref{Phi}]
\be
\label{all3phase}
\Phi(t;\btau) = \omega_{0}\frac{t}{\To} + \omega_{1}\bigg(\frac{t}{\To}\bigg)^2 + \omega_2\bigg(\frac{t}{\To}\bigg)^3
+ \omega_3\bigg(\frac{t}{\To}\bigg)^4 + \alpha_{1}\,\mu_{1}(t) + \alpha_{2}\,\mu_{2}(t),
\ee
where $\btau=(\omega_{0},\omega_{1},\omega_2,\omega_3,\alpha_{1},\alpha_{2})$.
The 6-dimensional reduced Fisher matrix $\tilde{\Gamma}$
with elements $\tilde{\Gamma}_{ij}$ ($i,j=1,\ldots,6$) given in Eq.\ \eqref{Gamma}
and for the phase $\Phi$ defined in Eq.\ \eqref{all3phase} equals
\be
\label{all3mF}
\tilde{\Gamma}^{\text{all-sky}}_{s=3}(\chii) = \begin{pNiceMatrix}
\Block{4-4}{\tilde{\Gamma}^{\text{dir}}_{s=3}(\chii)}
&&&
& \tilde{\Gamma}_{15}^{\text{all-sky}}(\chii)
& \tilde{\Gamma}_{16}^{\text{all-sky}}(\chii)
\\[1ex]
&&&
& \tilde{\Gamma}_{25}^{\text{all-sky}}(\chii)
& \tilde{\Gamma}_{26}^{\text{all-sky}}(\chii)
\\[1ex]
&&&
& \tilde{\Gamma}_{35}^{\text{all-sky}}(\chii)
& \tilde{\Gamma}_{36}^{\text{all-sky}}(\chii)
\\[1ex]
&&&
& \tilde{\Gamma}_{45}^{\text{all-sky}}(\chii)
& \tilde{\Gamma}_{46}^{\text{all-sky}}(\chii)
\\[1ex]
\cdots
& \cdots
& \cdots
& \cdots
& \tilde{\Gamma}_{55}^{\text{all-sky}}(\chii)
& \tilde{\Gamma}_{56}^{\text{all-sky}}(\chii)
\\[1ex]
\cdots
& \cdots
& \cdots
& \cdots
& \cdots
& \tilde{\Gamma}_{66}^{\text{all-sky}}(\chii)
\end{pNiceMatrix}
\ee
where $\tilde{\Gamma}^{\text{dir}}_{s=3}$ is given in Eq.\ \eqref{dir3mF} and where
\bse
\begin{align}
\tilde{\Gamma}_{15}^{\text{all-sky}}(\chii) &= \dlangle \chi\mu_{1} \drangle,
\\[1ex]
\tilde{\Gamma}_{16}^{\text{all-sky}}(\chii) &=\dlangle \chi\mu_{2} \drangle,
\\[1ex]
\tilde{\Gamma}_{25}^{\text{all-sky}}(\chii) &= \avii{\Big(\chi^2+2\chii\chi-\frac{1}{12}\Big)\mu_{1}},
\\[1ex]
\tilde{\Gamma}_{26}^{\text{all-sky}}(\chii) &= \avii{\Big(\chi^2+2\chii\chi-\frac{1}{12}\Big)\mu_{2}},
\\[1ex]
\tilde{\Gamma}_{35}^{\text{all-sky}}(\chii) &= \avii{\Big(\chi^3+3\chii\chi^2+3\chii^2\chi-\frac{1}{4}\chii\Big)\mu_{1}},
\\[1ex]
\tilde{\Gamma}_{36}^{\text{all-sky}}(\chii) &= \avii{\Big(\chi^3+3\chii\chi^2+3\chii^2\chi-\frac{1}{4}\chii\Big)\mu_{2}},
\\[1ex]
\tilde{\Gamma}_{45}^{\text{all-sky}}(\chii) &= \avii{\Big(\chi^4+4\chii\chi^3+6\chii^2\chi^2+4\chii^3\chi-\frac{1}{2}\chii^2-\frac{1}{80}\Big)\mu_{1}},
\\[1ex]
\tilde{\Gamma}_{46}^{\text{all-sky}}(\chii) &= \avii{\Big(\chi^4+4\chii\chi^3+6\chii^2\chi^2+4\chii^3\chi-\frac{1}{2}\chii^2-\frac{1}{80}\Big)\mu_{2}},
\\[1ex]
\tilde{\Gamma}_{55}^{\text{all-sky}}(\chii) &= \dlangle \mu_{1}^2 \drangle-\dlangle\mu_{1}\drangle^2,
\\[1ex]
\tilde{\Gamma}_{56}^{\text{all-sky}}(\chii) &= \dlangle\mu_{1} \mu_{2}\drangle-\dlangle\mu_{1}\drangle\dlangle\mu_{2}\drangle
\\[1ex]
\tilde{\Gamma}_{66}^{\text{all-sky}}(\chii) &= \dlangle \mu_{2}^2\drangle -\dlangle\mu_{2}\drangle^2.
\end{align}
\ese
The elements of the matrix $\tilde{\Gamma}^{\text{all-sky}}_{s=3}$ depend on the parameter $\chii$
and also on the time-dependent functions $\mu_1$ and $\mu_2$ introduced in Eq.\ \eqref{Phi},
which are determined by the motion of the detector with respect to the SSB.

\subsection{Network of detectors}

In this subsection, we derive the reduced Fisher matrix that takes into account
the information gathered by a network of $N$ GW detectors.
To do this we employ the same simplified model of the GW signal
we used above for the computation of single-detector Fisher matrices.
Thus the signal related with the $I$-th ($I=1,\ldots,N$) detector has the form [see Eqs.\ \eqref{hPhi}]
\be
\label{DNh}
h_I(t;h_{0I},\Phi_{0I},\bxi)
= h_{0I}\,\sin\big(\Phi_I(t;\bxi)+\Phi_{0I}\big), \quad
I=1,\ldots,N,
\ee
where $h_{0I}$ is a constant amplitude and $\Phi_{0I}$ is a constant initial phase of the signal.
The phase parameters $\bxi$ are common for all detectors,
but the phase functions $\Phi_I$ are in general different for different detectors
(they are different for all-sky searches and identical for all detectors in the case of directed searches).
The log likelihood ratio for the network,
assuming that noises in different detectors are independent stochastic processes, reads:
\be
\label{DNlnL}
\ln\Lambda_{\text{net}}[\mathbf{x};\mathbf{A},\bxi] \cong 2\,\To \sum_{I=1}^N \frac{1}{S_{nI}(\fc)}\,
\Big(\langle x_I(t)\,h_I(t;\mathbf{A}_I,\bxi) \rangle -\frac{1}{2}\langle h_I(t;\mathbf{A}_I,\bxi)^{2}\rangle\Big).
\ee
Here the vector $\mathbf{x}$ collects all data streams, $\mathbf{x}=(x_1,\ldots,x_N)$,
and the vector $\mathbf{A}$ consists of amplitude parameters, $\mathbf{A}:=(\mathbf{A}_1,\ldots,\mathbf{A}_N)$,
where $\mathbf{A}_I:=(h_{0I},\Phi_{0I})$ ($I=1,\ldots,N$).
After introducing $h_{1I}:=h_{0I}\cos\Phi_{0I}$ and $h_{2I}:=h_{0I}\sin\Phi_{0I}$,
one easily solves equations
$\partial\ln\Lambda_{\text{net}}/\partial h_{1I}=0$, $\partial\ln\Lambda_{\text{net}}/\partial h_{2I}=0$
for maximum-likelihood estimators of $h_{1I}$ and $h_{2I}$;
the solution reads $\hat{h}_{1I}=2\langle x_I\sin\Phi_I\rangle$ and $\hat{h}_{2I}=2\langle x_I\cos\Phi_I\rangle$.
The $\F$-statistic for the network one gets by replacing in \eqref{DNlnL} the amplitude parameters $\mathbf{A}$
by their estimators $\hat{\mathbf{A}}$:
\be
\label{DNF}
\F_{\text{net}}[\mathbf{x};\bxi] := \ln\Lambda_{\text{net}}[\mathbf{x};\hat{\mathbf{A}},\bxi]
\cong 2\,\To \sum_{I=1}^N \frac{1}{S_{nI}(\fc)}\,
\Big(\langle x_I(t)\,\sin\Phi_I(t;\bxi) \rangle^2 + \langle x_I(t)\,\cos\Phi_I(t;\bxi) \rangle^2\Big).
\ee

Let us now consider the situation when the data streams $x_I$ in all detectors
contain the GW signals $h_I$ coming from the same astrophysical source.
The signals have amplitude parameters $\mathbf{A}_I'$ (different for different detectors)
and phase parameters $\bxi'$ (the same for all detectors):
\be
x_I(t) = n_I(t) + h_I(t;\mathbf{A}_I',\bxi').
\ee
One can then show that the expectation value of the network $\F$-statistic \eqref{DNF} reads
\begin{align}
\label{DNE1F}
\text{E}_1\{\F_{\text{net}}[\mathbf{x};\bxi]\}
\cong N + \frac{1}{2} \sum_{I=1}^N (\rho_I(h'_{0I}))^2 \,C_{0I}(\bxi,\bxi'),
\end{align}
where $\rho_I(h'_{0I})$ is the signal-to-noise ratio in the $I$-th detector,
\be
\label{rhoI}
\rho_I(h'_{0I}) \cong h'_{0I} \sqrt{\frac{\To}{S_{nI}(\fc)}},
\ee
and $C_{0I}(\bxi,\bxi')$ is the signal-free [i.e., defined for $x_I(t) = n_I(t)$]
autocovariance function of the $\F$-statistic computed for the $I$-th detector
[see Eq.\ \eqref{C02}]:
\be
\label{DNC0I}
C_{0I}(\bxi,\bxi') \cong \big\langle\sin\big[\Phi_I(t;\bxi)-\Phi_I(t;\bxi')\big]\big\rangle^{2}
+ \big\langle\cos\big[\Phi_I(t;\bxi)-\Phi_I(t;\bxi')\big]\big\rangle^{2}.
\ee
Making use of \eqref{rhoI}, Eq.\ \eqref{DNE1F} can be rewritten as
\be
\label{DNE1Fa}
\text{E}_1\{\F_{\text{net}}[\mathbf{x};\bxi]\}
\cong N + \frac{\To}{2} \sum_{I=1}^N \frac{(h'_{0I})^2}{S_{nI}(\fc)} \,C_{0I}(\bxi,\bxi').
\ee

One can define the signal-free auotcovariance function of the network $\F$-statistic \eqref{DNF} as
\be
\label{DNC0}
C^{\text{net}}_0(\bxi,\bxi') \coloneqq \mathrm{E}\big\{[\F_{\text{net}}[\mathbf{n};\bxi]-m^{\text{net}}_0(\bxi)]
[\F_{\text{net}}[\mathbf{n};\bxi']-m^{\text{net}}_0(\bxi')]\big\},
\ee
where $m^{\text{net}}_0$ is the signal-free expectation value of $\F_{\text{net}}$:
\be
\label{DNm0}
m^{\text{net}}_0(\bxi) \coloneqq \mathrm{E}\{\F_{\text{net}}[\mathbf{n};\bxi]\}.
\ee
We have found that $C^{\text{net}}_0(\bxi,\bxi')$
is just the sum of the auotcovariance functions of the $\F$-statistics computed for individual detectors,
\be
\label{DNC01}
C^{\text{net}}_0(\bxi,\bxi') \cong \sum_{I=1}^N C_{0I}(\bxi,\bxi').
\ee

In the $I$-th detector the phase $\Phi_I$ reads
(note that the functions $\mu_{1I}$ and $\mu_{2I}$ are different for different detectors):
\begin{align}
\label{DNPhiI}
\Phi_I(t;\bxi) = \sum_{k=0}^{s} \omega_k \Big(\frac{t}{\To}\Big)^{k+1}
+ \alpha_{1}\,\mu_{1I}(t) + \alpha_{2}\,\mu_{2I}(t).
\end{align}
$\Phi_I$ depends linearly on the parameters $\bxi$,
consequently the one-detector autocovariance functions $C_{0I}(\bxi,\bxi')$
depend on $\btau:=\bxi-\bxi'$ and can be written as
\be
\label{DNC02}
C_{0I}(\btau) \cong \langle\cos\Phi_I(t;\btau)\rangle^{2}
+ \langle\sin\Phi_I(t;\btau)\rangle^{2}.
\ee
Next we replace the functions $C_{0I}(\btau)$
by the approximation derived in Eq.\ \eqref{Ca}, so
\be
\label{DNCaI}
C_{0I}(\btau) \cong C_{\mathrm{a}I}(\btau) \coloneqq
1 - \sum_{k=1}^{s+3}\sum_{l=1}^{s+3} (\tilde{\Gamma}_I)_{kl}\,\tau_{k}\,\tau_{l},
\ee
where $\tilde{\Gamma}_I$ is the $(s+3)$-dimensional
\emph{reduced Fisher matrix} of the $I$th detector with elements equal to
\be
\label{DNGammaI}
(\tilde{\Gamma}_I)_{kl} := \Big\langle\frac{\partial\Phi_I}{\partial\tau_{k}}
\frac{\partial\Phi_I}{\partial\tau_{l}}\Big\rangle -
\Big\langle\frac{\partial\Phi_I}{\partial\tau_{k}}
\Big\rangle \Big\langle\frac{\partial\Phi_I}{\partial\tau_{l}}
\Big\rangle, \quad k,l=1,\ldots,s+3.
\ee

It is clear that the right-hand side of Eq.\ \eqref{DNE1Fa} can not be directly expressed
by the network autocovariance function $C^{\text{net}}_0$,
because of the way it depends on the individual amplitudes $h'_{0I}$ of the GW signals
and on noise spectral densities $S_{nI}(\fc)$ of particular detectors.
In the next step we replace in Eq.\ \eqref{DNE1Fa} all amplitudes $h'_{0I}$
by the same for all detectors averaged amplitude ${h'_0}^\text{av}$.
This procedure can be justified as follows.
The signal-to-noise ratio (SNR) $\rho_I(h'_{0I})$ is computed here
for the simplified model \eqref{DNh} of the GW signal
and it represents some effective or averaged SNR.
The SNR for non-simplified GW signal actually observed in the detector was studied in detail
in section III~C of \cite{98JKS}. The ``exact'' SNR $\rho_I$ for each individual detector
obviously depends on the observation time $\To$
and on the parameters of both the GW source and the detector.
It depends on right ascension $\alpha$ and declination $\delta$ of the GW source,
polarization angle $\psi$ of the wave, and inclination angle $\iota$,
i.e.\ the angle between the total angular momentum of the star and the direction from the star to the Earth;
it also depends on the position of the detector on Earth,
and on orientation of its arms with respect to local geographical directions;
finally it depends on the spectral density $S_{nI}(\fc)$ of the detector's noise.
One can average the squared of the SNR with respect to the angles $\alpha$, $\delta$, $\psi$, and $\iota$.
The result is given in Eqs.\ (92) and (93) in \cite{98JKS} and it has the form
\be
\langle\rho_I^2\rangle_{\alpha,\delta,\psi,\iota} \cong \frac{(h_0^\text{av})^2\To}{S_{nI}(\fc)},
\ee
where $h_0^\text{av}$ does not depend on the position of the detector on Earth
and orientation of its arms, so it is the same for all detectors.

After replacing in Eq.\ \eqref{DNE1Fa} all amplitudes $h'_{0I}$
by the averaged amplitude ${h'_0}^\text{av}$one gets
\begin{align}
\label{DNE1F2}
\text{E}_1\{\F_{\text{net}}[\mathbf{x};\bxi]\}
\cong N + \frac{1}{2}\To({h'_0}^\text{av})^2 \sum_{I=1}^N \frac{1}{S_{nI}(\fc)}\,C_{0I}(\bxi,\bxi').
\end{align}
Making use of the approximation \eqref{DNCaI}, the above equation becomes
($\btau=\bxi-\bxi'$ and we employ matrix notation):
\be
\label{DNE1F3}
\text{E}_1\{\F_{\text{net}}[\mathbf{x};\bxi]\} \cong N + \frac{1}{2}\To({h'_0}^\text{av})^2
\sum_{I=1}^N \frac{1}{S_{nI}(\fc)}\big(1 - \btau^\mathsf{T} \cdot \tilde{\Gamma}_I \cdot \btau\big).
\ee
It is convenient to introduce
the harmonic mean of the detectors' noise spectral densities
(evaluated at the same central frequency $\fc$),
\be
S_{n\text{h}}(\fc) := N\Big(\sum_{I=1}^N \frac{1}{S_{nI}(\fc)}\Big)^{-1}.
\ee
Making use of the above definition Eq.\ \eqref{DNE1F3} can be rewritten as
\be
\label{DNE1F4}
\text{E}_1\{\F_{\text{net}}[\mathbf{x};\bxi]\} \cong N + \frac{1}{2}N\frac{\To({h'_0}^\text{av})^2}{S_{n\text{h}}(\fc)}
\Bigg(1 - \btau^\mathsf{T} \cdot \bigg(\frac{S_{n\text{h}}(\fc)}{N}\sum_{I=1}^N \frac{1}{S_{nI}(\fc)}\tilde{\Gamma}_I\bigg) \cdot \btau\Bigg).
\ee
This equation suggests that the reduced Fisher matrix for the network of detectors can be defined as
\be
\label{DNGamma}
\tilde{\Gamma}_{\text{net}} \coloneqq \frac{S_{n\text{h}}(\fc)}{N}\sum_{I=1}^N \frac{1}{S_{nI}(\fc)}\tilde{\Gamma}_I.
\ee
Let us observe that if the reduced Fisher matrices $\tilde{\Gamma}_I$ for individual detectors are identical for all detectors
(this is the case for directed searches, see sections \ref{SubsecRFmD3} and \ref{SubsecRFmD5} above),
i.e.\ $\tilde{\Gamma}_I=\tilde{\Gamma}$ for $I=1,\ldots,N$,
then the reduced Fisher matrix for the network equals the one-detector Fisher matrix, $\tilde{\Gamma}_{\text{net}}=\tilde{\Gamma}$.

\section{Construction of banks of templates}
\label{secBofT}

\subsection{Covering problem}

Construction of a bank of templates begins with the fixing of the minimum value $\cmin$
of the autocovariance function $C_0(\bxi,\bxi')$ of the signal-free $\F$-statistic.
One then looks for such a grid (i.e.\ a discrete set) of points in the space spanned by the parameters $\bxi$,
that for any GW signal with parameters $\bxi'$ that may be present in the data,
there exists a grid node with parameters $\bxi$ such that $C_0(\bxi,\bxi')\ge\cmin$.
We employ the approximation $C_0(\bxi,\bxi') \cong \Ca(\bxi,\bxi')$,
so by virtue of Eq.\ \eqref{Ca} this inequality can be written as
\be
\label{bank4}
\sum_{k,l=1}^{s+3}\tilde{\Gamma}_{kl}\,(\xi_k-\xi'_k)\,(\xi_l-\xi'_l)
\le 1 - \cmin.
\ee
For the fixed $\bxi$ and $0<\cmin<1$, Eq.\ \eqref{bank4} defines an \emph{hyperellipsoid}
in $(s+3)$-dimensional Euclidean space
with the center at $\bxi$ and with the size determined by $\cmin$.
The Euclidean volume of this hyperellipsoid reads\footnote{
We use in this work the factorial symbol $n!$ also when $n$ is not a nonnegative integer,
i.e.\ we employ the definition $n!:=\Gamma(n+1)$, where $\Gamma$ denotes the gamma function.}
(here $d=s+3$)
\be
V_d = \frac{\pi^{d/2}(1 - \cmin)^{d/2}}
{(d/2)!\,\sqrt{\det\tilde{\Gamma}}}.
\ee

For the fixed (and bounded) region of the parameter space,
we want to find such a grid fulfilling the requirement \eqref{bank4},
which consists of possibly smallest number of points.
Therefore the problem of finding the optimal grid is a \emph{covering} problem,
i.e.\ the problem to cover the $d$-dimensional Euclidean space $\mathbb{R}^d$
by the smallest number of \emph{identical} hyperellipsoids.
It is convenient to replace the problem of finding the optimal covering of space by identical hyperellipsoids
by the problem of finding the optimal covering of space by identical \emph{hyperspheres of unit radius},
we do this below in section \ref{HE2HS}.
A thorough presentation of the problem of covering $d$-dimensional Euclidean space $\mathbb{R}^d$
with identical hyperspheres (with not necessarily a unit radius)
can be found in Chap.\ 2 of Ref.\ \cite{CS99}.
Let us now recall the basic notions related to this problem.

We consider only $d$-dimensional \emph{lattice} coverings,
i.e.\ sets of points which are linear combinations
with integer coefficients of some basis vectors $(\mathbf{P}_1,\dots,\mathbf{P}_d)$,
$\mathbf{P}_i\in\mathbb{R}^d$ (for $i=1,\ldots,d$).
Let the coordinates of the basis vectors be:
$\mathbf{P}_1=(P_{11},\ldots,P_{1d})$, \ldots, $\mathbf{P}_d=(P_{d1},\ldots,P_{dd})$.
The matrix whose rows are the coordinates of vectors $\mathbf{P}_i$,
\be
\mathsf{M} = \left(
\begin{array}{cccc}
P_{11} & P_{12} & \cdots & P_{1d} \\
P_{21} & P_{22} & \cdots & P_{2d} \\
\vdots & \vdots & \ddots & \vdots \\
P_{d1} & P_{d2} & \cdots & P_{dd}
\end{array}
\right).
\ee
is called a \emph{generator matrix} for the lattice.
For each lattice one can define its \emph{fundamental region},
which when repeated fills the space with one lattice point in each copy.
An example of this is a \emph{fundamental parallelotope} collecting the points of the form
$\lambda_1 \mathbf{P}_1 + \ldots + \lambda_d \mathbf{P}_d$,
where $0 \le \lambda_1 < 1$, \ldots, $0 \le \lambda_d < 1$.
The quality of a covering is given by the \emph{covering thickness} $\Theta$
defined as the average number of hyperspheres that contain a point in the space.
For lattice coverings $\Theta$ is the ratio (volume of one hypersphere)/(volume of fundamental region).
Thickness of lattice covering of $\mathbb{R}^d$ with identical hyperspheres of radius $R$ reads
\be
\label{rho}
\Theta = \frac{\pi^{d/2}R^{d}}{(d/2)!\,|\det\mathsf{M}|},
\ee
where $\mathsf{M}$ is the generator matrix for the lattice.

For any discrete collection of points $\mathcal{Q}=\{\mathbf{Q}_1,\mathbf{Q}_2,\ldots\}\subset\mathbb{R}^d$
one defines its \emph{covering radius} $R$, as the least upper bound
for the distance from any point $\mathbf{x}\in\mathbb{R}^d$ to the closest point of $\mathcal{Q}$,
$R:=\sup_{\mathbf{x}\in\mathbb{R}^d} \inf_{\mathbf{Q}_i\in\mathcal{Q}} |\mathbf{x}-\mathbf{Q}_i|.$
The identical hyperspheres of radius $R$ centered at the points of $\mathcal{Q}$ will cover $\mathbb{R}^d$,
and no hyperspheres of radius smaller than $R$ will cover it.



\subsubsection{Constraints}

We consider searches with very large number of grid points in the parameter space,
so the time needed to compute the $\F$-statistic for all grid nodes is long
and we want to speed up this computation by employing the fast Fourier transform (FFT) algorithm.
As the FFT algorithm computes the values of the discrete Fourier transform (DFT) of a time series
for a certain set of discrete frequencies (the Fourier frequencies),
it will be possible to use the FFT algorithm
if the frequency coordinates of grid points
will all coincide with the Fourier frequencies.
We will construct grids which fulfill this requirement.

If the data form a sequence of $N$ samples
$x_u$, $u=1,\ldots,N$,
with the sampling-in-time period $\Delta{t}$ (so $N\Delta{t}=\To$),
then the frequency resolution of the DFT is $\Delta{f}=1/(N\Delta{t})$
and the resolution of the parameter $\omega_0$
[defined in Eq.\ \eqref{omegak}] is
\be
\label{Domega0}
\Delta{\omega_0} = 2\pi\To \Delta{f} = 2\pi.
\ee
One can improve the frequency resolution of the DFT by adding $N_{\text{FFT}}-N$ zeros to the $N$ samples
(so one takes the Fourier transform of $N_{\text{FFT}}$ data points),
then the frequency resolution \eqref{Domega0} changes to\footnote{
One can also diminish the frequency resolution of the DFT
by folding of the data; see Appendix in Ref.\ \cite{2011PJP} for more details.}
\be
\label{FFTwiaz1}
\Delta\omega_{0} = 2\pi\,\frac{N}{N_{\text{FFT}}}.
\ee
We thus need such grid that all nodes can be arranged along straight lines parallel to the $\omega_{0}$-axis
and the distance between neighboring nodes along these lines must be equal to $\Delta\omega_0$.
To fulfil this condition it is enough to require that the first basis vector has components\footnote{
From now on we treat all $d$-vectors as \emph{column} $d\times1$ matrices
and we use matrix notation with superscript ``$\mathsf{T}$'' denoting matrix transposition
and ``$\cdot$'' denoting matrix multiplication.}
\be
\label{wiazP1}
\mathbf{P}_1 = (\Delta\omega_0,0,\ldots,0)^\mathsf{T}.
\ee

In the case of all-sky searches there is another important constraint
related to resampling of the data to the so called \emph{barycentric time}
(see e.g.\ section III D in \cite{98JKS}, section V A in \cite{ABJPK10}, \cite{PSDB2010},
and section 6.2 in \cite{2014LSC/VC-CQGb}).
Because numerically accurate resampling is computationally demanding,
it is crucial to do resampling only once per sky position for all spindown values.
Let us arrange the parameters $\bxi$ in all-sky searches in such a way
that the parameters $\alpha_1 $ and $\alpha_2$ (related to the location of the source in the sky)
are the last two components of the vector $\bxi$, see Eq.\ \eqref{vecxi}.
Then, to fulfill the above-mentioned requirement,
it is enough to require that the last two components of the second basis vector vanish:
\be
\label{wiazP2}
\mathbf{P}_2 = (P_{21},P_{22},\ldots,P_{2 d-2},0,0)^\mathsf{T}.
\ee

\subsubsection{Replacing hyperellipsoid-coverings by hypersphere-coverings}
\label{HE2HS}

We replace the problem of finding the optimal covering of space by identical hyperellipsoids
by the problem of finding the optimal covering of space by identical hyperspheres of \textit{unit radius}.
Let us denote the space of original parameters $\btau$ by $\Omega$
($\Omega\subset\mathbb{R}^d$, $\btau\in\Omega$)
and the space of the transformed parameters $\btau'$ by $\Omega'$
($\Omega'\subset\mathbb{R}^d$, $\btau'\in\Omega'$).
We consider a linear transformation from $\Omega$ to $\Omega'$ defined be some matrix $\mathsf{F}$,
\be
\label{HE2HS01}
\btau' = \mathsf{F} \cdot \btau.
\ee
We require that \eqref{HE2HS01} transforms the hyperellipsoid defined by the constant value $\cmin$ of the autocovariance function
into the hypersphere of unit radius.
The hyperellipsoid with its center at the origin of the coordinate system
is determined in the $\Omega$ space by the equation [see Eq.\ \eqref{bank4}]
\be
\label{HE2HS02}
\btau^\top \cdot \tilde{\Gamma}\cdot \btau = 1 - \cmin,
\ee
whereas the hypersphere of unit radius in the $\Omega'$ space
is described by the relation ${\btau'}^\top\cdot\btau' = 1$,
which, after making use of \eqref{HE2HS01}, can be written as
\be
\label{HE2HS03}
\btau^\top\cdot(\mathsf{F}^\top\cdot\mathsf{F})\cdot\btau = 1.
\ee
Equations \eqref{HE2HS02} and \eqref{HE2HS03} describe the same hyperellipsoid if and only if
\be
\label{HE2HS04}
\mathsf{F}^\top \cdot \mathsf{F} = \frac{1}{R_\text{av}^2}\tilde{\Gamma},
\ee
where $R_\text{av}\coloneqq\sqrt{1-\cmin}$ is the average radius of the hyperellipsoid \eqref{HE2HS02}.
The elements of the matrix $\tilde{\Gamma}$ depend on the parameter $\chii$,
therefore Eq.\ \eqref{HE2HS04} implies that the elements of the matrix $\mathsf{F}$
depend on the parameters $\chii$ and $\cmin$.
Additionally, in the case of all-sky searches, the elements of $\mathsf{F}$ depend,
through the evaluation of time averages needed to compute the elements of the Fisher matrix [see section \ref{SubsecRFmAS}],
on the position vector of the detector with respect to the SSB during observational interval.

All considered in section \ref{SecRFmatrices} reduced Fisher matrices $\tilde{\Gamma}$ are symmetric and strictly positive definite,
i.e.\ $\btau^\top\cdot\tilde{\Gamma}\cdot\btau>0$ for any $\btau\ne\mathbf{0}$,
therefore Eq.\ \eqref{HE2HS04} can be interpreted as the \emph{Cholesky decomposition} of the matrix $\tilde{\Gamma}/R_\text{av}^2$.
Consequently there exists the unique \emph{upper triangular matrix} $\mathsf{F}$,
with strictly positive diagonal elements, fulfilling Eq.\ \eqref{HE2HS04}.
Making use of the explicit formulas for the matrices $\tilde{\Gamma}$ from section \ref{SecRFmatrices},
one easily checks that for all matrices $\tilde{\Gamma}$ considered in section \ref{SecRFmatrices}
the $(1,1)$ element of the corresponding matrix $\mathsf{F}$ equals
\be
\label{F11}
F_{11} = \frac{1}{2\sqrt{3(1-\cmin)}}.
\ee

Now we transform the two constraints \eqref{wiazP1} and \eqref{wiazP2}
into the $\Omega'$ space.
We start with transforming the first basis vector $\mathbf{P}_1$
[with $\Omega$-space components given in Eq.\ \eqref{wiazP1}],
\be
\mathbf{P}'_1 \coloneqq \mathsf{F}\cdot\mathbf{P}_1
= \mathsf{F}\cdot(\Delta\omega_0,0,\ldots,0)^\top.
\ee
Because the matrix $\mathsf{F}$ is upper triangular,
this leads to
\be
\label{wiaz1}
\mathbf{P}'_1 = (\Delta\omega_{0}',0,\ldots,0)^\top,
\ee
where, by means of Eq.\ \eqref{F11},
the length $\Delta\omega_{0}'$ of the vector $\mathbf{P}'_1$ equals
\be
\label{omegavsomega}
\Delta\omega_0' = \frac{\Delta\omega_0}{2\sqrt{3(1-\cmin)}}.
\ee
The second basis vector $\mathbf{P}_2$ has $\Omega$-space components given in \eqref{wiazP2}.
Its image in the in $\Omega'$ space also has its two last components equal to zero:
\be
\label{wiaz2}
\mathbf{P}'_2 \coloneqq \mathsf{F}\cdot\mathbf{P}_2
= (P'_{21},P'_{22},\ldots,P'_{2d-2},0,0)^\top.
\ee

When the basis vectors of the grid in $\Omega'$ space are found,
one transform them into $\Omega$ space by means of the inverse matrix $\mathsf{F}^{-1}$.
If $\mathsf{S}'$ is the generator matrix of the grid in $\Omega'$ space,
then the generator matrix $\mathsf{S}$ of the corresponding grid in $\Omega$ space can be computed as
\be
\label{odo}
\mathsf{S} = \mathsf{S}' \cdot (\mathsf{F}^{-1})^\top.
\ee
All generator matrices $\mathsf{S}'$ we construct below are lower triangular.
It implies that also generator matrices $\mathsf{S}$ are lower triangular
(note that the matrix $\mathsf{F}^{-1}$ is upper triangular so its transpose is lower triangular).

\subsection{Construction of grids}
\label{S1S2construction}

In this subsection we describe constructions of grids,
which fulfil the constraints given by Eqs.\ \eqref{wiaz1}--\eqref{omegavsomega} and \eqref{wiaz2}.
All grids we build in $\Omega'$ space,
they only depend on the value of the parameter $\Delta\omega_0'$,
that is, through Eq.\ \eqref{omegavsomega},
they depend on the values of the search parameters $\Delta\omega_{0}$ and $\cmin$.
The grids we construct are based on simple deformations
of the $d$-dimensional lattice coverings $A_d^{\star}$ of space $\mathbb{R}^d$
by \emph{hyperspheres of unit radius}.
We construct two families of grids, called $S_1$ and $S_2$.
The construction of the $S_1$ grids boils down to squeezing the $A_d^{\star}$ lattice in one appropriately selected direction,
while the construction of the $S_2$ grids is more complex,
although it also consists in deforming the fundamental region of the $A_d^{\star}$ lattice.
The grids $S_2$ have mostly smaller than the grids $S_1$ thicknesses for smaller values of the parameter $\Delta\omega_0'$.

\subsubsection{$S_1$ grids}
\label{subsecS1}

In this subsection we describe the construction of the family $S_1$ of grids.
We start by finding such a vector $\mathbf{q}$ of the lattice $A_d^\star$
joining the origin of the space $\mathbb{R}^d$ with a lattice node,
the Euclidean length $|\mathbf{q}|$ of which is as close as possible to, but not less than $\Delta\omega_0'$
[$\Delta\omega_0'$ is the length of the vector $\mathbf{P}'_1$, see Eq.\ \eqref{wiaz1}].
The smaller the difference, the better---the lattice thickness after its deformation, i.e.\ compression,
will be closer to the thickness of the lattice $A_d^\star$.
In the limiting case $\Delta\omega_0'=|\mathbf{q}|$,
the grid $S_1$ coincides with the lattice $A_d^\star$.
We will illustrate the procedure of construction of generator matrix for the $S_1$ grid in more detail
in the ($d=5$)-dimensional case.
We denote coordinates of any point in $\mathbb{R}^5$ by $(\omega_0',\omega_1',\omega_2',\alpha_1',\alpha_2')$.

In 5 dimensions, the generator matrix of the lattice $A_5^\star$ with covering radius equal to 1
can be calculated from Eq.\ \eqref{Md1}, using Eqs.\ \eqref{M6} and \eqref{R2toR6}.
The explicit form of this matrix reads
\be
\label{M5'}
\mathsf{M}_{5,1} = \left(\begin{array}{ccccc}
2\sqrt{\frac{3}{7}} & 0 & 0 & 0 & 0\\
-\frac{2\sqrt{\frac{3}{7}}}{5} & \frac{12\sqrt{\frac{2}{7}}}{5} & 0 & 0 & 0\\
-\frac{2\sqrt{\frac{3}{7}}}{5} & -\frac{3\sqrt{\frac{2}{7}}}{5}  & 3\sqrt{\frac{6}{35}} & 0 & 0\\
-\frac{2\sqrt{\frac{3}{7}}}{5} & -\frac{3\sqrt{\frac{2}{7}}}{5}  & -\sqrt{\frac{6}{35}} & 4\sqrt{\frac{3}{35}} & 0\\
-\frac{2\sqrt{\frac{3}{7}}}{5} & -\frac{3\sqrt{\frac{2}{7}}}{5}  & -\sqrt{\frac{6}{35}} & -2\sqrt{\frac{3}{35}} & \sqrt{\frac{6}{35}}
\end{array}\right).
\ee
To find the vector $\mathbf{q}$, it is convenient to modify the generator matrix of the lattice $A_d^\star$
in such a way as to search only the half-space defined by nonnegative values of the coordinate $\omega_0'$.
In the 5-dimensional case the generator matrix of $A_5^\star$ that provides this reads
\be
\label{M5''}
\mathsf{M}^1_{5,1} = \left(
\begin{array}{ccccc}
2\sqrt{\frac{3}{7}} & 0 & 0 & 0 & 0\\
\frac{2\sqrt{\frac{3}{7}}}{5} & \frac{12\sqrt{\frac{2}{7}}}{5} & 0 & 0 & 0\\
\frac{2\sqrt{\frac{3}{7}}}{5} & -\frac{3\sqrt{\frac{2}{7}}}{5} & 3\sqrt{\frac{6}{35}} & 0 & 0\\
\frac{2\sqrt{\frac{3}{7}}}{5} & -\frac{3\sqrt{\frac{2}{7}}}{5} & -\sqrt{\frac{6}{35}} & 4\sqrt{\frac{3}{35}} & 0\\
\frac{2\sqrt{\frac{3}{7}}}{5} & -\frac{3\sqrt{\frac{2}{7}}}{5} & -\sqrt{\frac{6}{35}} & -2\sqrt{\frac{3}{35}} & \sqrt{\frac{6}{35}}
\end{array}
\right).
\ee
The matrix \eqref{M5''} can be obtained from the matrix $\mathsf{M}_{5,1}$
by making reflection of the lattice generated by $\mathsf{M}_{5,1}$ with respect to the hyperplane $\omega_0'=0$
and then replacing the first basis vector with the opposite vector
(let us recall that components of the basis vectors form the rows of the generator matrix).

We look for the vector $\mathbf{q}$ in the form of a linear combination
of the basis vectors $\mathbf{v}_a$ of the lattice $A_d^\star$:
\be
\label{wektor2}
\mathbf{q}_{i_{1},\ldots,i_{d}} \coloneqq \sum_{a=1}^{d}\,i_a\mathbf{v}_a,
\ee
where $i_{1},\ldots,i_{d}$ are integers.
Further, we require that the chosen vector $\mathbf{q}$ together with some $d-1$ vectors taken out of $d$ basis vectors $\mathbf{v}_a$,
form a new basis for the lattice $A_d^\star$.
This way we construct a new generator matrix with the components of the vector $\mathbf{q}$ forming,
say, the first raw of the matrix.
Next we will rotate the lattice several times
to make the vector $\mathbf{q}$ parallel to the $\omega_0'$-axis.
We will display below the necessary rotation matrices in the 5-dimensional case.

Let us denote the components of the chosen vector $\mathbf{q}$
(of the length $|\mathbf{q}|$ closest to $\Delta\omega_0'$) in 5 dimensions by
\be
\label{wektorq}
\mathbf{q} = (q_1, q_2, q_3, q_4, q_5)^\top,
\quad
|\mathbf{q}| = \sqrt{\sum_{a=1}^5 q_a^2}.
\ee
The new generator matrix of $A_5^\star$ looks like this
\be
\label{V0}
\mathsf{M}^2_{5,1} = \left(\begin{array}{ccccc}
q_1 & q_2 & q_3 & q_4 & q_5 \\
p_{1} & p_{2} & p_{3} & p_{4} & p_{5} \\
r_{1} & r_{2} & r_{3} & r_{4} & r_{5} \\
s_{1} & s_{2} & s_{3} & s_{4} & s_{5} \\
t_{1} & t_{2} & t_{3} & t_{4} & t_{5} 
\end{array} \right),
\ee
where $(p_1,\ldots,p_5)^\top$, $(r_1,\ldots,r_5)^\top$, $(s_1,\ldots,s_5)^\top$, and $(t_1,\ldots,t_5)^\top$
are the components of four out of five basis vectors $\{\mathbf{v}_1,\ldots,\mathbf{v}_5\}$
selected in such a way as to create together with the vector $\mathbf{q}$
a new basis for the $A_5^\star$ lattice.
We then first rotate the lattice in the $(\alpha_1',\alpha_2')$ plane by the angle $\beta_1$,
using the rotation matrix $\mathsf{R}_1(\beta_1)$:
\be
\label{ob}
\mathsf{R}_1(\beta_1) \coloneqq \left(\begin{array}{ccccc}
1 & 0 & 0 & 0 & 0  \\
0 & 1 & 0 & 0 & 0 \\
0 & 0 & 1 & 0 & 0 \\
0 & 0 & 0 & \cos\beta_1 & -\sin\beta_1 \\
0 & 0 & 0 & \sin\beta_1 & \cos\beta_1
\end{array}\right),
\ee
where the angle $\beta_1$ satisfies the equations
\be
\sin\beta_1 = \frac{q_5}{\sqrt{q_4^2+q_5^2}},
\quad
\cos\beta_1 = \frac{q_4}{\sqrt{q_4^2+q_5^2}}.
\ee
We obtain the new generator matrix $\mathsf{M}^3_{5,1}$,
\be
\label{V1}
\mathsf{M}^3_{5,1} = \mathsf{M}^2_{5,1}\cdot\mathsf{R}_1(\beta_1)
= \left(\begin{array}{ccccc}
q_1 & q_2 & q_3 & \sqrt{q_4^2+q_5^2} & 0 \\
p_{1} & p_{2} & p_{3} & p_{4}' & p_{5}' \\
r_{1} & r_{2} & r_{3} & r_{4}' & r_{5}' \\
s_{1} & s_{2} & s_{3} & s_{4}' & s_{5}' \\
t_{1} & t_{2} & t_{3} & t_{4}' & t_{5}' 
\end{array} \right).
\ee
Next, we perform rotation in the $(\omega_2',\alpha_1')$ plane
using the rotation matrix
\be
\label{og}
\mathsf{R}_2(\beta_2) \coloneqq \left(\begin{array}{ccccc}
 1 & 0 & 0 & 0 & 0 \\
 0 & 1 & 0 & 0 & 0 \\
 0 & 0 & \cos\beta_2 & -\sin\beta_2 & 0 \\
 0 & 0 & \sin\beta_2 & \cos\beta_2 & 0 \\
 0 & 0 & 0 & 0 & 1 \\
\end{array}\right),
\ee
where the angle $\beta_2$ satisfies the equations
\be
\sin\beta_2 = \frac{\sqrt{q_4^2+q_5^2}}{\sqrt{q_3^2+q_4^2+q_5^2}},
\quad
\cos\beta_2 = \frac{q_3}{\sqrt{q_3^2+q_4^2+q_5^2}}.
\ee
We get
\be
\label{V2}
\mathsf{M}^4_{5,1} = \mathsf{M}^3_{5,1}\cdot\mathsf{R}_2(\beta_2)
= \left(\begin{array}{ccccc}
q_1 & q_2 & \sqrt{q_3^2+q_4^2+q_5^2} & 0 & 0 \\
p_{1} & p_{2} & p_{3}' & p_{4}'' & p_{5}' \\
r_{1} & r_{2} & r_{3}' & r_{4}'' & r_{5}' \\
s_{1} & s_{2} & s_{3}' & s_{4}'' & s_{5}' \\
t_{1} & t_{2} & t_{3}' & t_{4}'' & t_{5}' 
\end{array} \right).
\ee  
In the next step we make rotation in the $(\omega_1',\omega_2')$ plane
by the angle $\beta_3$ such, that
\be
\sin\beta_3 = \frac{\sqrt{q_3^2+q_4^2+q_5^2}}{\sqrt{q_2^2+q_3^2+q_4^2+q_5^2}},
\quad
\cos\beta_3 = \frac{q_2}{\sqrt{q_2^2+q_3^2+q_4^2+q_5^2}}.
\ee
This rotation is described by the matrix
\be
\label{od}
\mathsf{R}_3(\beta_3) \coloneqq \left(\begin{array}{ccccc}
1 & 0 & 0 & 0 & 0 \\
0 & \cos\beta_3 & -\sin\beta_3 & 0 & 0 \\
0 & \sin\beta_3 & \cos\beta_3 & 0 & 0 \\
0 & 0 & 0 & 1 & 0 \\
0 & 0 & 0 & 0 & 1 
\end{array}\right),
\ee
and the resulting generator matrix reads
\be
\label{V3}
\mathsf{M}^5_{5,1} = \mathsf{M}^4_{5,1}\cdot\mathsf{R}_3(\beta_3)
= \left(\begin{array}{ccccc}
q_1 & \sqrt{q_2^2+q_3^2+q_4^2+q_5^2} & 0 & 0 & 0 \\
p_{1} & p_{2}' & p_{3}'' & p_{4}'' & p_{5}' \\
r_{1} & r_{2}' & r_{3}'' & r_{4}'' & r_{5}' \\
s_{1} & s_{2}' & s_{3}'' & s_{4}'' & s_{5}' \\
t_{1} & t_{2}' & t_{3}'' & t_{4}'' & t_{5}' 
\end{array} \right).
\ee  
Finally, the lattice generated by the matrix \eqref{V3} we rotate in the $(\omega_0',\omega_1')$ plane
using the rotation matrix
\be
\label{ot}
\mathsf{R}_4(\beta_4) \coloneqq \left(\begin{array}{ccccc}
\cos\beta_4 & -\sin\beta_4& 0 & 0 & 0 \\
\sin\beta_4 & \cos\beta_4& 0 & 0 & 0 \\
0 & 0 & 1 & 0 & 0 \\
0 & 0 & 0 & 1 & 0 \\
0 & 0 & 0 & 0 & 1 
\end{array}\right),
\ee
where the angle $\beta_4$ is defined through equations
\be
\sin\beta_4 = \frac{\sqrt{q_2^2+q_3^2+q_4^2+q_5^2}}{|\mathbf{q}|},\quad
\cos\beta_4 = \frac{q_1}{|\mathbf{q}|}.
\ee
This way we obtain the generator matrix of the lattice $A_5^\star$ in the form
\be
\label{V4}
\mathsf{M}^6_{5,1} = \mathsf{M}^5_{5,1}\cdot\mathsf{R}_4(\beta_4)
= \left(\begin{array}{ccccc}
|\mathbf{q}| & 0 & 0 & 0 & 0 \\
p_{1}' & p_{2}'' & p_{3}'' & p_{4}'' & p_{5}' \\
r_{1}' & r_{2}'' & r_{3}'' & r_{4}'' & r_{5}' \\
s_{1}' & s_{2}'' & s_{3}'' & s_{4}'' & s_{5}' \\
t_{1}' & t_{2}'' & t_{3}'' & t_{4}'' & t_{5}' 
\end{array} \right).
\ee  

Next we apply the above described procedure to the basis vector from the second raw of the matrix \eqref{V4}.
We rotate the lattice generated by the matrix $\mathsf{M}^6_{5,1}$
successively in the $(\alpha_1',\alpha_2')$, $(\omega_2',\alpha_1')$, and $(\omega_1',\omega_2')$ planes,
to replace the components $p_{3}''$, $p_{4}''$, and $p_{5}''$ by zero.
We can continue this procedure for the remaining basis vectors and we are always able to obtain a lower-triangular generator matrix,
regardless of the dimension $d$ of the grid.
After applying the above described procedure to all basis vectors of the lattice $A_5^\star$ in 5 dimensions,
we obtain the generator matrix of the form
\be
\label{V5}
\mathsf{M}^7_{5,1} = \left(\begin{array}{ccccc}
|\mathbf{q}| & 0 & 0 & 0 & 0 \\
p_{1} & p_{2} & 0 & 0 & 0 \\
r_{1} & r_{2} & r_{3} & 0 & 0 \\
s_{1} & s_{2} & s_{3} & s_{4} & 0 \\
t_{1} & t_{2} & t_{3} & t_{4} & t_{5}
\end{array}\right).
\ee

To get the generator matrix for the grid $S_1$,
we have to deform the lattice $A_{5}^{\star}$ generated by the matrix \eqref{V5}
by squeezing it in the direction of the $\omega_{0}'$-axis.
We squeeze the lattice by the factor $s\le1$ equal to
\be
\label{q1}
s \coloneqq \frac{1}{|\mathbf{q}|}\Delta\omega_{0}'.
\ee
The generator matrix for the grid $S_{1}$ finally equals
\be
\label{C1b}
\mathsf{S}_1 = \mathsf{M}^7_{5,1}\cdot\mathsf{S},
\ee
where the diagonal matrix $\mathsf{S}$ reads
\be
\label{Q1}
\mathsf{S} = \left(\begin{array}{ccccc}
 s & 0 & 0 & 0 & 0 \\
 0 & 1 & 0 & 0 & 0 \\
 0 & 0 & 1 & 0 & 0 \\
 0 & 0 & 0 & 1 & 0 \\
 0 & 0 & 0 & 0 & 1 
\end{array}\right).
\ee
The first two rows of the matrix $\mathsf{S}_{1}$ are components of the basis vectors
which fulfill the two constraints \eqref{wiaz1} and \eqref{wiaz2}.

By means of Eqs.\ \eqref{rho} and \eqref{rhoAstar} one easily computes the thickness of the grid $S_1$
generated by the matrix $\mathsf{S}_1$. It equals
\be
\label{rhoS1b1}
\Theta_{S_{1}} = \frac{245\pi^2}{3888}\sqrt{\frac{35}{3}}\frac{|\mathbf{q}|}{\Delta\omega_0'},
\ee
or, expressing $\Delta\omega_0'$ by search parameters $\Delta\omega_0$ and $\cmin$,
\be
\label{rhoS1b2}
\Theta_{S_{1}} = \frac{245\pi^2}{3888}\sqrt{\frac{35}{3}}\frac{|\mathbf{q}|\sqrt{1-\cmin}}{\Delta\omega_0}
\cong 2.1243 \frac{|\mathbf{q}|\sqrt{1-\cmin}}{\Delta\omega_0}.
\ee

\subsubsection{$S_2$ grids}
\label{subsecS2}

For completeness, we consider below in section \ref{secDiscussion} another family of grids,
designed in Refs.\ \cite{2011PJP,2015PJ}, and named (in \cite{2015PJ}) $S_2$.
They were constructed only in dimensions 2 (in \cite{2011PJP}) and 4 (in \cite{2015PJ})
and they are, as in the case of $S_1$ grids, some deformations of the lattice coverings $A_2^*$ and $A_4^*$, respectively.
The grid $S_2$ was constructed in dimension 4 in section 4.4 of Ref.\ \cite{2015PJ},
whereas the construction of $S_2$ in 2 dimensions, based on the results of section III~C of Ref.\ \cite{2011PJP},
is presented in detail in appendix \ref{AS2-2D} of this work.

\section{Discussion}
\label{secDiscussion}

\sloppy Constructions of the grids $S_1$ and $S_2$, described in section \ref{S1S2construction},
we have implemented in the computer program \texttt{GridsGenerator},
freely available on \texttt{GitHub} at \url{https://github.com/apisarski/gridgen}.
Most of the results discussed in the current section were obtained by means of this program.
With this program it is possible to compute generator matrices of grids
suitable for both direct and all-sky searches,
and with different numbers of spindown parameters included.
The program uses Fisher matrices derived in section \ref{SecRFmatrices},
in each case finds the matrix $\mathsf{F}$ [see Eq.\ \eqref{HE2HS01}] which translates the problem of covering by hyperellipsoids
into the problem of covering by unit hyperspheres,
then constructs the generator matrix of covering by unit hyperspheres fulfilling the constraints \eqref{wiaz1} and \eqref{wiaz2},
and finally, using the matrix $\mathsf{F}^{-1}$,
translate the generator matrix into the original parameter space [see Eq.\ \eqref{odo}].

\begin{figure*}
\begin{center}
\includegraphics[scale=0.25]{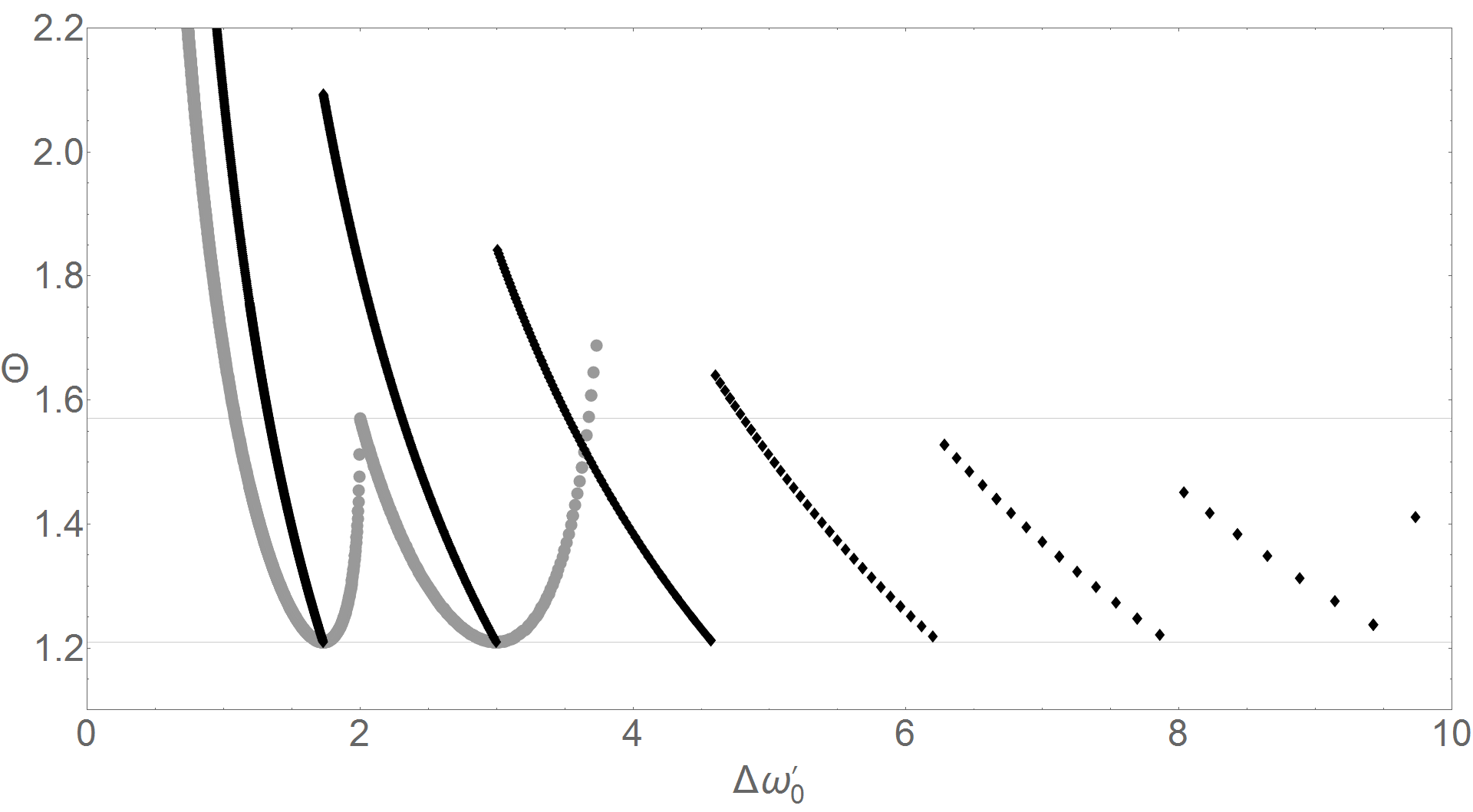}
\caption{\label{S1S2-2D}
Covering thicknesses $\Theta$ of the 2-dimensional grids $S_1$ (black diamonds) and $S_2$ (grey circles)
as functions of the quantity $\Delta\omega_0'$.
The upper horizontal line corresponds to the thickness of the 2-dimensional cubic lattice $\mathbb{Z}^2$ (it equals $\cong1.5708$)
and the lower horizontal line denotes the thickness of the optimal hexagonal lattice $A_2^\star$ (it equals $\cong1.2092$).
The grids can be applied for directed searches with one spindown parameter included.}
\end{center}
\end{figure*}

\begin{figure*}
\begin{center}
\includegraphics[scale=0.25]{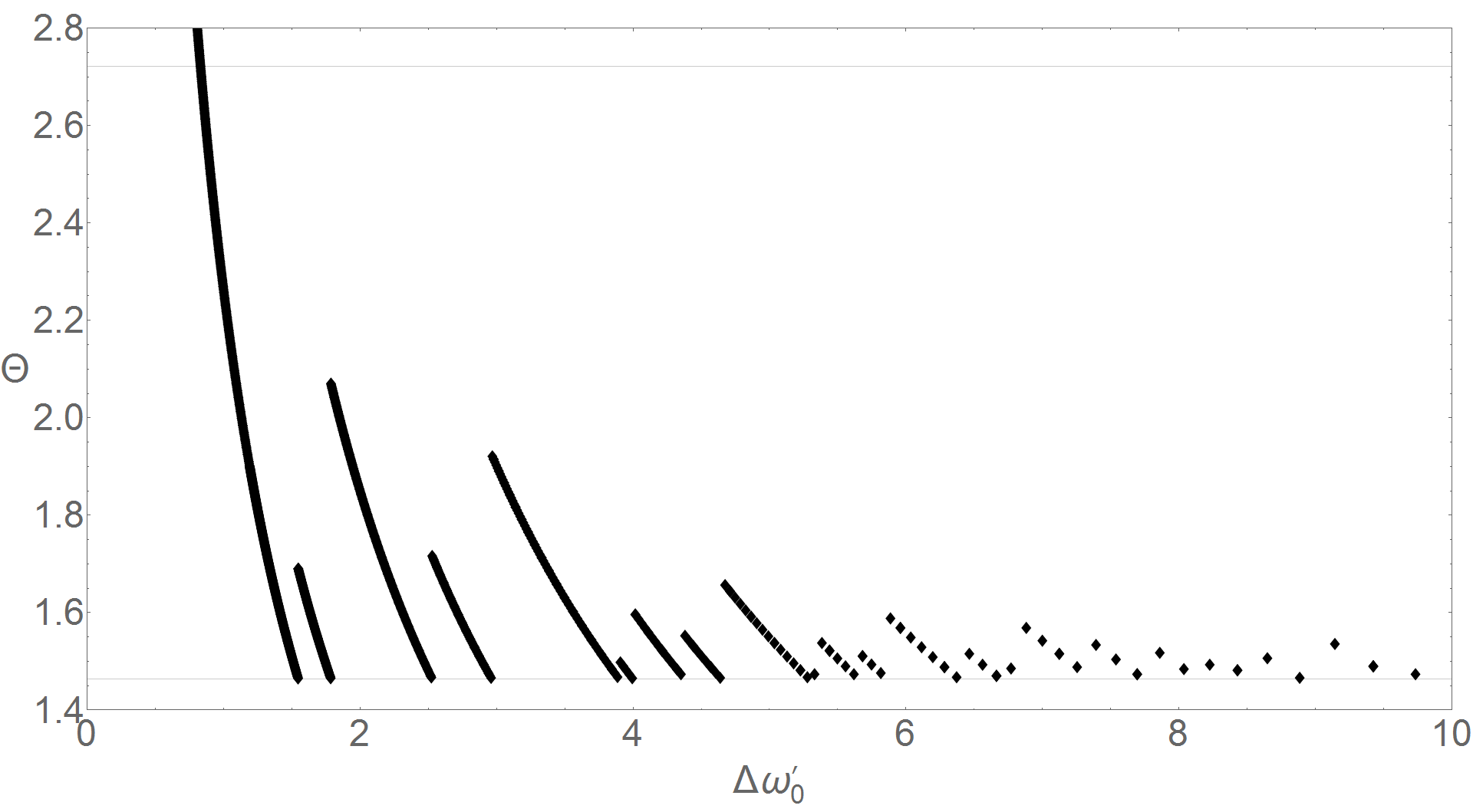}
\caption{\label{S1-3D}
Covering thicknesses $\Theta$ of the 3-dimensional grid $S_1$ (black diamonds)
as a function of the quantity $\Delta\omega_0'$.
The upper horizontal line corresponds to the thickness of the 3-dimensional cubic lattice $\mathbb{Z}^3$ (it equals $\cong2.7207$)
and the lower horizontal line denotes the thickness of the optimal lattice $A_3^\star$ (it equals $\cong1.4635$).
The grids can be applied for directed searches with two spindown parameters included
and for all-sky searches with no spindown parameters included.}
\end{center}
\end{figure*}

\begin{figure*}
\begin{center}
\includegraphics[scale=0.25]{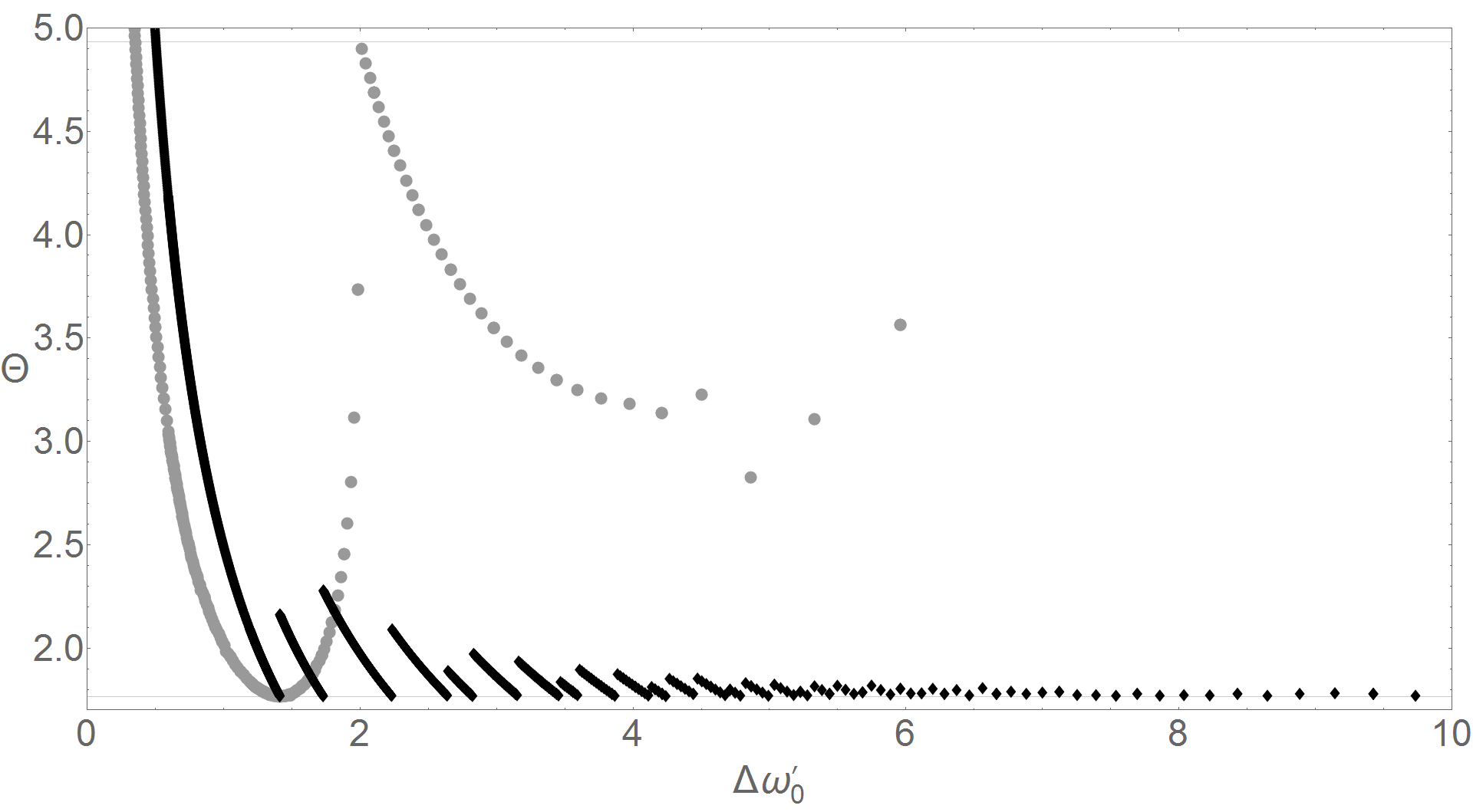}
\caption{\label{S1S2-4D}
Covering thicknesses $\Theta$ of the 4-dimensional grids $S_1$ (black diamonds) and $S_2$ (grey circles)
as functions of the quantity $\Delta\omega_0'$.
The upper horizontal line corresponds to the thickness of the 4-dimensional hypercubic lattice $\mathbb{Z}^4$ (it equals $\cong4.9348$)
and the lower horizontal line denotes the thickness of the optimal lattice $A_4^\star$ (it equals $\cong1.7655$).
The grids can be applied for directed searches with three spindown parameters included
and for all-sky searches with one spindown parameter included.
The graph made with black diamonds and the part of the graph made with grey circles for $\Delta\omega_0'\gtrsim0.5$
are identical to the corresponding graphs in Fig.\ 4 from work \cite{2015PJ} (where they were presented as possible to be used only
in all-sky searches with one spindown parameter included).}
\end{center}
\end{figure*}

\begin{figure*}
\begin{center}
\includegraphics[scale=0.25]{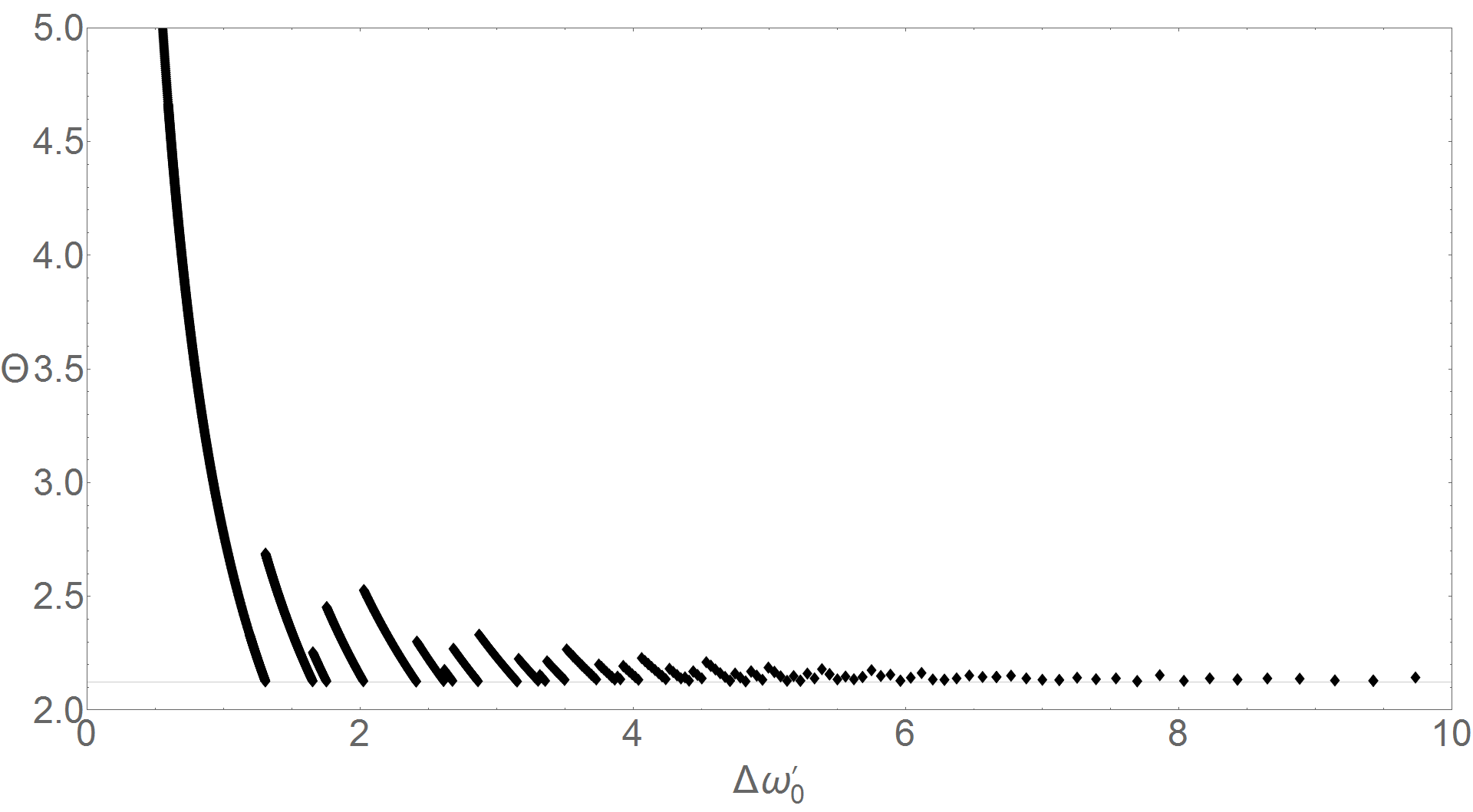}
\caption{\label{S1-5D}
Covering thicknesses $\Theta$ of the 5-dimensional grids $S_1$ (black diamonds)
as a function of the quantity $\Delta\omega_0'$.
The lower horizontal line denotes the thickness of the optimal lattice $A_5^\star$ (it equals $\cong2.1243$).
The thickness of the 5-dimensional hypercubic lattice $\mathbb{Z}^5$ equals $\cong9.1955$,
so it is beyond the displayed range of the vertical axis.
The grids can be applied for directed searches with four spindown parameters included
and for all-sky searches with two spindown parameters included.}
\end{center}
\end{figure*}

\begin{figure*}
\begin{center}
\includegraphics[scale=0.25]{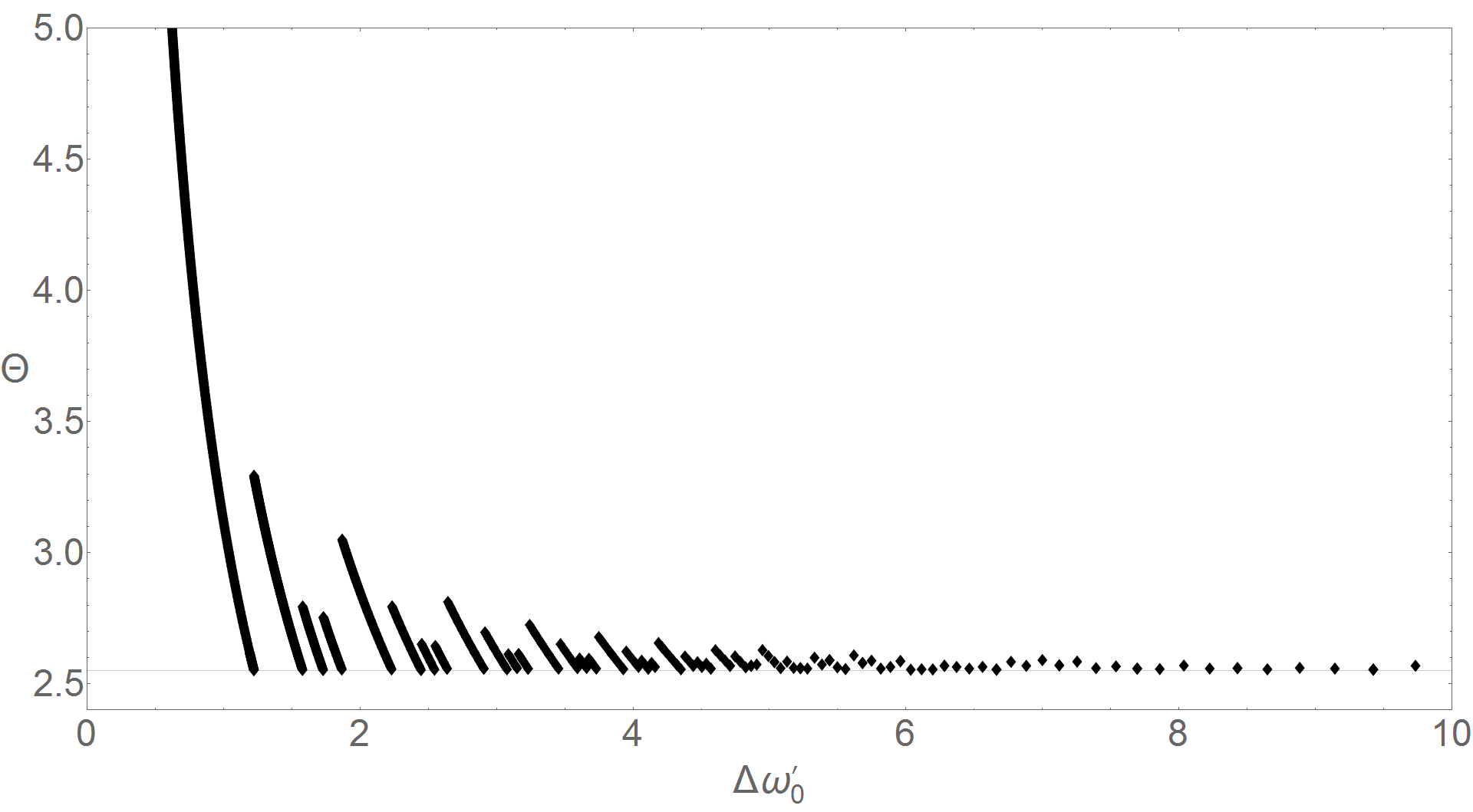}
\caption{\label{S1-6D}
Covering thickness $\Theta$ of the 6-dimensional grids $S_1$ (black diamonds)
as a function of the quantity $\Delta\omega_0'$.
The lower horizontal line denotes the thickness of the lattice $A_6^\star$ (it equals $\cong2.5511$).
The thickness of the 6-dimensional hypercubic lattice $\mathbb{Z}^6$ equals $\cong17.4410$,
so it is beyond the displayed range of the vertical axis.
The grids can be applied for directed searches with five spindown parameters included
and for all-sky searches with three spindown parameters included.}
\end{center}
\end{figure*}

We have constructed the grids $S_1$ and $S_2$ for different values of the parameter $\Delta\omega_0'$
and in different dimensions:
the grids $S_1$ in all dimensions from 2 to 6 and the grids $S_2$ only in dimensions 2 and 4.
In figures \ref{S1S2-2D}--\ref{S1-6D} we depict covering thicknesses $\Theta$ of the grids $S_1$ and $S_2$
as functions of the quantity $\Delta\omega_0'$.
Figure \ref{S1S2-2D} shows 2-dimensional grids $S_1$ and $S_2$
which can be applied for directed searches with one spindown parameter included.
3-dimensional grids $S_1$ are suitable for directed searches with two spindown parameters included
and for all-sky searches with no spindown parameters included,
they are depicted in figure \ref{S1-3D}.
Figure \ref{S1S2-4D} shows 4-dimensional grids $S_1$ and $S_2$
applicable for directed searches with three spindown parameters included
and all-sky searches with one spindown parameter included\footnote{
For 4-dimensional case, at the moment of writing this article, the released version of the program
\texttt{GridsGenerator} is able to generate grids $S_1$ and $S_2$ for all-sky searches
and only grid $S_1$ for directed searches.}.
Figures \ref{S1-5D} and \ref{S1-6D}\footnote{The results shown in figure \ref{S1-6D} are generated using
a version of the program \texttt{GridsGenerator} not yet released at the time of writing this article.}
depict the grids $S_1$ in 5 and 6 dimensions, respectively.
5-dimensional grids can be applied for directed searches with four spindown parameters included
and for all-sky searches with two spindown parameters included, whereas 6-dimensional grids
are suitable for directed searches with five spindown parameters included
and for all-sky searches with three spindown parameters included.

In figures \ref{S1S2-2D}--\ref{S1-6D}, we have marked by horizontal lines the value of the thickness of the $A_d^\star$ lattice,
and in figures \ref{S1S2-2D}--\ref{S1S2-4D}, we have also marked
the thickness of the corresponding hypercubic $\mathbb{Z}^d$ lattice\footnote{
The analytical formula for the thickness of the $\mathbb{Z}^d$ lattice reads
$(\pi d/4)^{d/2}/(d/2)!$.}
(the thicknesses of the $\mathbb{Z}^5$ and $\mathbb{Z}^6$ hypercubic lattices
are not shown in figures \ref{S1-5D} and \ref{S1-6D}, respectively,
as their values are beyond the displayed range of the vertical axis).
The $A_d^\star$ lattices are the thinnest known lattices in dimensions $2\le d\le5$,
and they are close to the thinnest known lattices in dimensions $6\le d\le15$
(compare table 2.1 in \cite{CS99} with table 2 in \cite{2008DSSV}).

Allen in Ref.\ \cite{2021Allen} claims,
that the thinnest template bank gives the most constraining upper ``strict'' limit
(strict means that it applies at every point in parameter space),
but it does not maximize the expected number of detections under assumptions
that both the number of templates and the total volume of the parameter space covered by them, are fixed.
The expected number of detections is maximized by grids which are ``optimal quantizers'' \cite{2021Allen}.
For small mismatches the optimal quantizer minimizes the scale-invariant second moment $G$ of the lattice
[defined in Eq.\ (5.8) of \cite{2021Allen}],
which measures the average squared distance from the nearest template (or the average expected mismatch).
Let us note that the lattices $A_2^\star$ and $A_3^\star$ are also the best lattice quantizers,
and from table I of \cite{2021Allen} it follows, that the lattices $A_4^\star$--$A_6^\star$ used by us
have $G$ only 1.2\%--3.0\% larger than the best known quantizers in dimensions 4--6
(the same table shows that $A_d^\star$ lattices in dimensions 7--15 have $G$ larger by 0.6\%--7.8\%
compared to the best known quantizers).
At this point, it should also be emphasized that the grids $S_1$ and $S_2$ constructed in this work
meet the constraints related to the possibility of using the FFT algorithm
and fast resampling to the barycentric time.
There is a huge difference, in terms of the computational cost needed to calculate the $\mathcal{F}$-statistic for all grid's nodes,
between the situation when the applied data analysis pipeline uses the fulfillment of these constraints and when it does not.
This means that for a given total computational cost, the constraint-compliant pipeline
will enable searching a much larger portion of the parameter space.
Considering this, and noting that our grids,
which are generally small deformations of the $A_d^\star$ lattices,
have, firstly, near-optimal thicknesses\footnote{\label{O1footnote}
As an example, let us note that the time-domain $\mathcal{F}$-statistic-based all-sky search of O1 data
used 4-dimensional $S_1$ grids with thicknesses 1.7796
(for low-frequency part of the search, see section IV D of \cite{2017LSC/VC-PRDa})
and 1.7677 (for higher-frequency search, see section VI of \cite{2018LSC/VC-PRDa}),
which are only $\sim$0.8\% and $\sim$0.1\% larger than the thickness 1.7655 of the optimal $A_4^\star$ lattice.},
and, secondly, have values of $G$ also not much different from the best known values\footnote{
Note, referring to footnote \ref{O1footnote},
that the 4-dimensional lattice with the smallest known $G$ is $D_4$ (see table I in \cite{2021Allen}),
which has $G$ only 1.2\% smaller than the $A_4^\star$ lattice. At the same time,
the thickness of the $D_4$ is $\sim$40\% greater than the thickness of the $A_4^\star$.},
using these grids for data analysis is very well justified.

The grids constructed in this work are, and we expect that they will be employed
in the first part of continuous-wave searches using the time-domain $\F$-statistic pipeline \cite{2014LSC/VC-CQGb}
(the second part is the search for coincidences between the candidates obtained in the first part).
The first part is the coherent search of time-domain but narrowband in frequency data segments.
In all-sky $\F$-statistic-based searches of O1--O3 data \cite{2017LSC/VC-PRDa,2018LSC/VC-PRDa,2019LSC/VC-PRDa,2022LSC/VC-arXiv},
the data were split into frequency bands 0.25 Hz long.
For each band, the data were inverse Fourier transformed to get a time series.
In O1--O3 searches, data segments of 2 to 24 sidereal days were analyzed (with longer segments for lower frequency ranges),
and after performing the search $10^{10}$--$10^{12}$ candidates were obtained, which were then screened for coincidences.
In all these searches, to compute the $\F$-statistic, an exact model of the GW signal with two spindowns was used,
and to construct banks of templates, the approximate linear model with also two spindowns was employed.
The study performed in Ref.\ \cite{JK99} showed that these models were adequate
for lengths of observational intervals and frequency/first frequency derivative ranges used in the searches of O1--O3 data.
However, it can be expected that longer data segments will be coherently analyzed in future searches,
making it necessary to use signal models and template banks with two or more spindowns included.
Such banks of templates are constructed in the current work.

\begin{acknowledgments}

The work was supported in part by the Polish NCN Grant
No.\ UMO-2017/26/M/ST9/00978.
We would like to thank Andrzej Królak for reading the manuscript and commenting on it.

\end{acknowledgments}

\appendix

\section{$A_d^\star$ lattices}
\label{appendixAdstar}

The generator matrix $\mathsf{M}_d$ for the lattice $A_d^{\star}$ can be obtained from the formula
\be
\label{AnStar1}
\mathsf{M}_d \cdot \mathsf{M}_d^{\top} = \text{Gram}(d),
\ee
where $\text{Gram}(d)$ is called a \textit{Gram matrix} for the lattice,
and it equals (see \cite{CS99}, p.\ 115):
\be
\label{Gram}
\text{Gram}(d) := \left(\begin{array}{ccccc}
d & -1 & -1 & \cdots & -1 \\
-1 & d & -1 & \cdots & -1 \\
\vdots & \vdots & \vdots & \ddots & \vdots \\
-1 & -1 & -1 & \cdots & d
\end{array}\right).
\ee
The generator matrix $\mathsf{M}_d$ defined by means of Eqs.\ \eqref{AnStar1} and \eqref{Gram} is a lower triangular matrix
(so $\mathsf{M}_d^{\top}$ is an upper triangular matrix).
The covering radius of the lattice  $A_d^\star$ generated by the matrix $\mathsf{M}_d$ reads
(see Theorem 7 and Eq.\ (51) on p.\ 474 of \cite{CS99})
\be
\label{Rn}
R_d = \frac{1}{2\sqrt{3}} \sqrt{d\left(d+2\right)}.
\ee
To obtain the $A_d^{\star}$ lattice with covering radius equal to one,
it is enough to divide the generator matrix $\mathsf{M}_d$ by $R_d$.
Thus the generator matrix $\mathsf{M}_{d,1}$ for the lattice $A_d^{\star}$ with unit covering radius is equal
\be
\label{Md1}
\mathsf{M}_{d,1} = \frac{1}{R_d}\mathsf{M}_d.
\ee

The generator matrices $\mathsf{M}_d$ for $d\in\{2,3,4,5,6\}$,
computed by means of Cholesky decomposition algorithm using Eqs.\ \eqref{AnStar1} and \eqref{Gram}, read
\begin{subequations}
\begin{align}
\label{M2}
\mathsf{M}_2 &= \left(\begin{array}{cc}
 \sqrt{2} & 0 \\
 -\frac{1}{\sqrt{2}} & \sqrt{\frac{3}{2}} \\
\end{array}\right),
\\[2ex]
\label{M3}
\mathsf{M}_3 &= \left(\begin{array}{ccc}
 \sqrt{3} & 0 & 0 \\
 -\frac{1}{\sqrt{3}} & 2 \sqrt{\frac{2}{3}} & 0 \\
 -\frac{1}{\sqrt{3}} & -\sqrt{\frac{2}{3}} & \sqrt{2} \\
\end{array}\right),
\\[2ex]
\label{M4}
\mathsf{M}_4 &= \left(\begin{array}{cccc}
2 & 0 & 0 & 0 \\
-\frac{1}{2} & \frac{\sqrt{15}}{2} & 0 & 0 \\
-\frac{1}{2} & -\frac{\sqrt{\frac{5}{3}}}{2} & \sqrt{\frac{10}{3}} & 0 \\
-\frac{1}{2} & -\frac{\sqrt{\frac{5}{3}}}{2} & -\sqrt{\frac{5}{6}} & \sqrt{\frac{5}{2}}
\end{array}\right),
\\[2ex]
\label{M5}
\mathsf{M}_5 &= \left(\begin{array}{ccccc}
\sqrt{5} & 0 & 0 & 0 & 0\\
-\frac{1}{\sqrt{5}} & 2\sqrt{\frac{6}{5}} & 0 & 0 & 0\\
-\frac{1}{\sqrt{5}} & -\sqrt{\frac{3}{10}} & \frac{3}{\sqrt{2}} & 0 & 0\\
-\frac{1}{\sqrt{5}} & -\sqrt{\frac{3}{10}} & -\frac{1}{\sqrt{2}} &  2 & 0\\
-\frac{1}{\sqrt{5}} & -\sqrt{\frac{3}{10}} & -\frac{1}{\sqrt{2}} & -1 & \sqrt{3}
\end{array}\right),
\\[2ex]
\label{M6}
\mathsf{M}_6 &= \left(\begin{array}{cccccc}
\sqrt{6} & 0 & 0 & 0 & 0 & 0 \\
-\frac{1}{\sqrt{6}} & \sqrt{\frac{35}{6}} & 0 & 0 & 0 & 0 \\
-\frac{1}{\sqrt{6}} & -\sqrt{\frac{7}{30}} & 2 \sqrt{\frac{7}{5}} & 0 & 0 & 0 \\
-\frac{1}{\sqrt{6}} & -\sqrt{\frac{7}{30}} & -\frac{\sqrt{\frac{7}{5}}}{2} & \frac{\sqrt{21}}{2} & 0 & 0 \\
-\frac{1}{\sqrt{6}} & -\sqrt{\frac{7}{30}} & -\frac{\sqrt{\frac{7}{5}}}{2} & -\frac{\sqrt{\frac{7}{3}}}{2} & \sqrt{\frac{14}{3}} & 0 \\
-\frac{1}{\sqrt{6}} & -\sqrt{\frac{7}{30}} & -\frac{\sqrt{\frac{7}{5}}}{2} & -\frac{\sqrt{\frac{7}{3}}}{2} & -\sqrt{\frac{7}{6}} & \sqrt{\frac{7}{2}} \\
\end{array}\right).
\end{align}
\end{subequations}
The covering radii of the lattice coverings generated by the matrices $\mathsf{M}_2$--$\mathsf{M}_6$,
according to Eq.\ \eqref{Rn}, equal
\be
\label{R2toR6}
R_2 = \sqrt{\frac{2}{3}},
\quad
R_3 = \frac{1}{2}\sqrt{5},
\quad
R_4 = \sqrt{2},
\quad
R_5 = \frac{1}{2}\sqrt{\frac{35}{3}},
\quad
R_6 = 2.
\ee

The thickness of the lattice $A_d^{\star}$ can be computed from the formula
(see \cite{CS99}, p.\ 36):
\be
\label{rhoAstar}
\Theta_d = \frac{\left(\frac{\pi}{12}d(d+2)\right)^{d/2}}{(d/2)!\,(d+1)^{(d-1)/2}}.
\ee
The thicknesses of the lattices $A_2^{\star}$--$A_6^{\star}$ equal
\begin{align}
\Theta_2 &= \frac{2 \pi }{3 \sqrt{3}} \cong 1.2092,
\quad
\Theta_3 = \frac{5 \sqrt{5} \pi }{24} \cong 1.4635,
\quad
\Theta_4 = \frac{2 \pi ^2}{5 \sqrt{5}} \cong 1.7655,
\nonumber\\
\Theta_5 &= \frac{245\sqrt{\frac{35}{3}}\pi^2}{3888} \cong 2.1243,
\quad
\Theta_6 = \frac{32 \pi ^3}{147 \sqrt{7}} \cong 2.5511.
\end{align}

\section{Construction of the 2-dimensional grid $S_2$}
\label{AS2-2D}

The construction of the grid $S_2$ is performed in the $(\omega'_0,\omega'_1)$-plane
and it depends on whether the parameter $\Delta\omega'_0$ is smaller or larger than 2.
The construction of the grid $S_2$ for the case $\Delta\omega'_0 < 2$
is identical to the construction of the grid $G_{2,0}$ done in Ref.\ \cite{2011PJP}.
For completeness, this construction is repeated below in section \ref{AS2-2D-1}
using the notation from this work.
In section \ref{AS2-2D-2}, we describe the construction of the grid $S_2$
in the case $\Delta\omega'_0\in[2,4)$, not considered in Ref.\ \cite{2011PJP}.

\subsection{Grid $S_2$ for $\Delta\omega'_0 < 2$}
\label{AS2-2D-1}

\begin{figure}
\begin{center}
\includegraphics[scale=0.25]{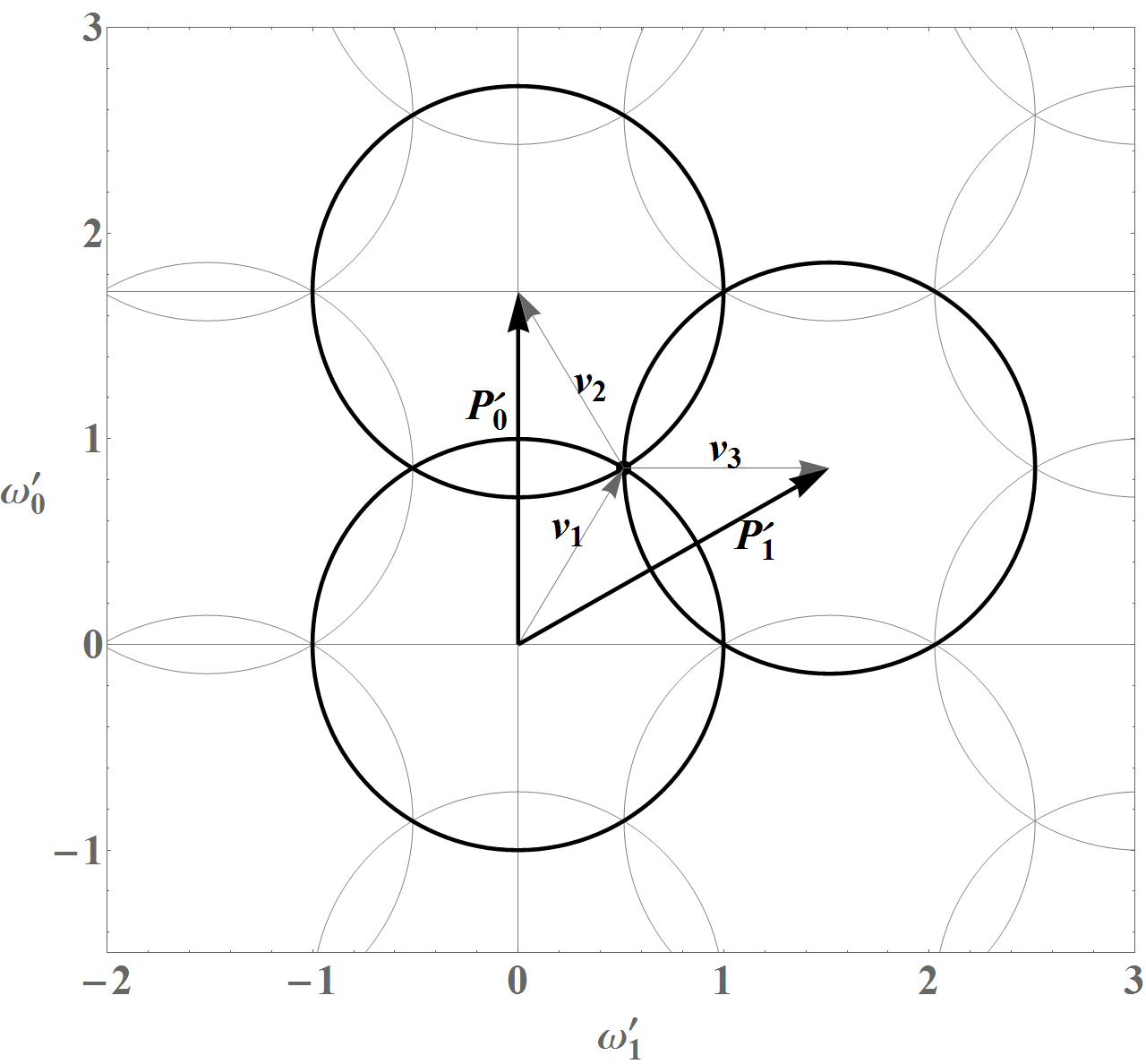}
\caption{\label{3circles1}
Construction of the 2-dimensional grid $S_2$ for $\Delta\omega'_0 < 2$.}
\end{center}
\end{figure}

We start from constructing three circles of unit radii which all cross each other at some point,
see figure \ref{3circles1}.
We build then the three vectors $\mathbf{v}_{1}$, $\mathbf{v}_{2}$, and $\mathbf{v}_{3}$,
all of unit length ($|\mathbf{v}_{1}|=|\mathbf{v}_{2}|=|\mathbf{v}_{3}|=1$).
The vector $\mathbf{v}_{1}$ begins at the centre of one of the circles
and ends at the point which is common to all three circles,
the vectors $\mathbf{v}_{2}$ and $\mathbf{v}_{3}$ are constructed in a similar way,
see figure \ref{3circles1}.
The components of the vectors $\mathbf{v}_{1}$, $\mathbf{v}_{2}$, and $\mathbf{v}_{3}$ are
\be
\label{218}
\mathbf{v}_{1}=(p,q), \quad
\mathbf{v}_{2}=(p,-q), \quad
\mathbf{v}_{3}=(0,1),
\ee
where $p$ and $q$ are positive numbers fulfilling the condition
\be
\label{213}
p^2+q^2=1.
\ee
From figure \ref{3circles1} it can be seen that
\be
\mathbf{P}'_0 = \mathbf{v}_1 + \mathbf{v}_2 = (2p,0).
\ee
Because the length of the vector $\mathbf{P}'_0$ has to be equal to $\Delta\omega'_0$,
we get
\be
\label{216}
p=\frac{1}{2}\Delta\omega'_0,
\ee
and
\be
\label{p0}
\mathbf{P}'_0 = (\Delta\omega'_0,0).
\ee
Making then use of Eqs.\ \eqref{213}, \eqref{216} and the equality
(see figure \ref{3circles1})
\be
\label{217}
\mathbf{P}'_1 = \mathbf{v}_{1}+\mathbf{v}_{3},
\ee
one easily obtains the components of the second basis vector $\mathbf{P}'_1$,
\be
\label{215}
\mathbf{P}'_1 = \Bigg(\frac{1}{2}\Delta\omega'_0, 1 + \sqrt{1-\Big(\frac{1}{2}\Delta\omega'_0\Big)^2}\Bigg).
\ee

\subsection{Grid $S_2$ for $\Delta\omega'_0\in[2,4)$}
\label{AS2-2D-2}

\begin{figure}
\begin{center}
\includegraphics[scale=0.25]{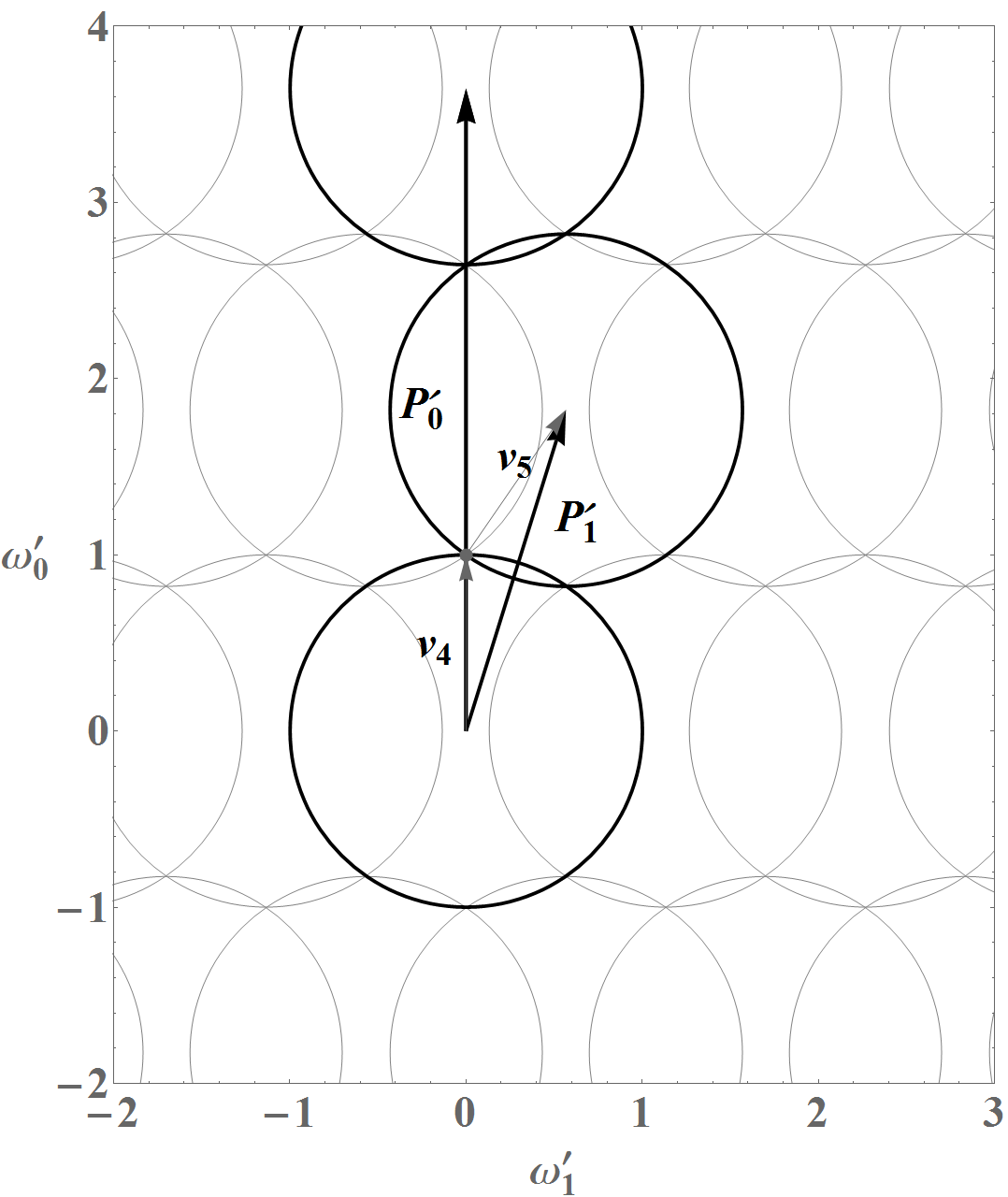}
\caption{\label{3circles2}
Construction of the 2-dimensional grid $S_2$ for $\Delta\omega'_0\in[2,4)$.}
\end{center}
\end{figure}

We start again from constructing three circles of unit radii, see figure \ref{3circles2}.
We build then the two vectors $\mathbf{v}_{4}$ and $\mathbf{v}_{5}$
of unit length ($|\mathbf{v}_{4}|=|\mathbf{v}_{5}|=1$):
the vector $\mathbf{v}_{4}$ begins at the centre of the bottom circle
and ends at the point which is common to the bottom circle and the vector $\mathbf{P}'_0$,
and the vector $\mathbf{v}_{5}$ begins at the point where the vector $\mathbf{v}_{4}$ ends
and ends at the point which belong to centre of the middle circle.
The components of the vectors $\mathbf{v}_{4}$ and $\mathbf{v}_{5}$ are
\be
\label{219}
\mathbf{v}_{4}=(1,0), \quad
\mathbf{v}_{5}=(p,q),
\ee
where $p$ and $q$ are positive numbers fulfilling the condition \eqref{213}.
From figure \ref{3circles2} it can easily be seen that
\be
\label{220}
|\mathbf{P}'_0| - 2 = 2p,
\ee
this leads to (remembering that $|\mathbf{P}'_0|=\Delta\omega'_0$)
\be
\label{221}
p = \frac{1}{2}\Delta\omega'_0 - 1.
\ee
Making then use of Eqs.\ \eqref{213}, \eqref{221} and the equality
(see figure \ref{3circles2})
\be
\label{227}
\mathbf{P}'_1 = \mathbf{v}_{4}+\mathbf{v}_{5},
\ee
one easily obtains the components of the second basis vector $\mathbf{P}'_1$,
\be
\label{225}
\mathbf{P}'_1 = \Bigg(\frac{1}{2}\Delta\omega'_0, \sqrt{\Delta\omega'_0\Big(1-\frac{1}{4}\Delta\omega'_0\Big)}\Bigg).
\ee

The above-described construction of the grid $S_2$ can be performed only if
$2 \leqslant |\mathbf{P}'_0| < 4$.
When $|\mathbf{P}'_0|=4$, the vectors $\mathbf{P}'_0$ and $\mathbf{P}'_1$ become parallel to each other.

\end{document}